\documentclass{article}
\usepackage{amssymb}
\usepackage{amsmath}
\usepackage{graphicx}
\newtheorem{defi}{Definition}[section]
\newtheorem{rem}{Remark}[section]
\newtheorem{thm}{Theorem}[section]
\newtheorem{lemma}[thm]{Lemma}
\newtheorem{cor}[thm]{Corollary}
\newtheorem{prop}[thm]{Proposition}
\def\rmi{{\rm i}}

\newcommand\hasira{{\mbox{\large\strut}}}

\newcommand\qed{\hfill\blacksquare\bigskip}
\newcommand\und{\mathrm{Und}}
\newcommand\ft{\mathrm{Ft}}
\newcommand\lam{\lambda}

\newcommand\hen[1]{\frac{\partial}{\partial {#1}}}
\newcommand\vect[1]{\mbox{\boldmath$#1$}}
\newcommand\bunsuu[1]{\frac{b_{#1}}{c_{#1}}}
\newcommand\BBB{\ensuremath{B^\circ}}
\newcommand\Zzz{\ensuremath{\vect{z}}^\circ}
\newcommand\zzz[1]{\ensuremath{{z_j}^\circ}}
\newcommand\ZZ{\ensuremath{\mathbb{Z}}}
\newcommand\CC{\ensuremath{\mathbb{C}}}

\newcommand\RR{\ensuremath{\mathbb{R}}}
\newcommand\NN{\ensuremath{\mathbb{N}}}

\newcommand\el{\ensuremath{\mathrm{El}}}
\newcommand\bbs[1]{\ensuremath{\mathcal{Z}_{#1}}}
\newcommand\lcm{\ensuremath{\mathop{\mathrm{LCM}}}}
\newcommand\ep{\varepsilon}
\newcommand\zet[1]{\left\vert {#1} \right\vert}
\newcommand\ud{\mbox{\,$-\!\triangleright$\,}}
\newcommand\udsim{\mbox{\,$\stackrel{\mathrm{u}}{\sim}$\,}}
\newcommand\proof{{\textbf {Proof.}}\ \,}
\begin{document}
\title{Ultradiscretization of the solution of
periodic Toda equation}
\author{Shinsuke Iwao and Tetsuji Tokihiro}
\maketitle
\begin{center}
Graduate School of Mathematical Sciences, 
The University of Tokyo\\ 
3-8-1
Komaba Meguro-ku, Tokyo  153-8914, Japan\\
iwao@ms.u-tokyo.ac.jp,toki@ms.u-tokyo.ac.jp
\end{center}
\begin{abstract}
A periodic box-ball system (pBBS)
is obtained by ultradiscretizing the periodic discrete Toda equation (pd Toda eq.).
We show
the relation
between a Young diagram of the pBBS and a spectral curve of the pd Toda eq..
The formula for
the fundamental cycle of the pBBS is obtained as a colloraly.
\end{abstract}

\section{Preface}

A cellular automaton (CA) is a discrete dynamical system which consists
of an array of 
a number of cells. Each cell allows for finitely many states which change into new
states in discrete time.
Usually the rule of time evolution with which the system is equipped is quite simple,
and CA are often investigated as simple models for natural or social 
phenomena. The box-ball system (BBS) is one type of CA, represented by
finitely many balls and countably many boxes arranged in a line.

In this paper, we study
a periodic box-ball system (pBBS), 
which is a BBS with a periodic boundary condition.
The pBBS is closely related to integrable nonlinear equations.
In fact, the pBBS has soliton-like solutions 
and a large number of conserved quantities. Moreover, the pBBS
 can be obtained from integrable equations by the method of 
`$\mbox{ultra}$discretization'. 

This relation gives us a new method to describe the behaviour of a pBBS.
If the initial-value problem of integrable equations related to the pBBS
is solvable by some analytical method, the initial-value problem 
of pBBS itself is also solvable,
as the solution of pBBS is
obtained from the solution of the integrable equations
by ultradiscretization.

The present paper is organized as follows. 
In section \ref{ne}, we introduce the definition of the pBBS and the pd Toda equation.
These two objects are connected each other through `ultradiscretization'.
We define the conserved quantities of these two systems
and state a main theorem (Theorem \ref{renkon})
which yields direct relation between the spectral curve and the Young diagram.
Section \ref{ultra} is spend to prove theorem \ref{renkon}.
In Section.\ref{usi} and \ref{tora}, we give the solution of the initial value problem
of pBBS and derive the
fundamental period for it, as a corollary of theorem \ref{renkon}.
\section{Periodic box-ball system and periodic discrete Toda equation}
\label{ne}
\subsection{pBBS}\label{pBBS}
\begin{figure}[htbp]
\begin{center}
\includegraphics[viewport=79 321 410 769, clip,
width=6cm,height=7.7cm]{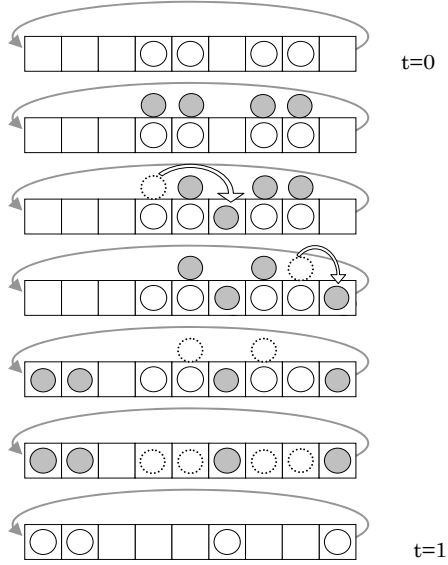}
\end{center}
\caption[Time evolution rule for pBBS]{Time evolution rule for pBBS}
\label{rule}
\end{figure}

Let us consider a one-dimensional array of $L$ boxes. Let $Q$ be
the total number of balls, such that $Q<L/2$.
Each of these boxes is either
empty or is filled with a ball. 
Since we are interested in the periodic case,
the $N$-th box is adjacent to the first box.
The time evolution of this system is: 
\begin{enumerate}
\def\labelenumi{(\theenumi)}
\def\theenumi{\roman{enumi}}
\item In each filled box, create a copy of the ball.
\item Move all copies once according to the following rules.
\item Choose one of the copies and move it to the nearest empty box
on the right of it.
\item Choose one of the remaining copies and move it to the nearest empty box
on the right of it.
\item Repeat the above procedure until all the copies have been moved.
\item Delete all the original balls.
\end{enumerate}
It is not difficult to confirm that
the resulting state does not depend on the choice of the copies.
This dynamical system is called the periodic box-ball system, or
pBBS. 
Figure \ref{rule} shows an example of
the pBBS and its time evolution pattern.
The last entry is considered to be adjacent to the first entry.
The pBBS is usually regarded as a dynamical system of a finite sequence
with periodic boundary condition.
Let us denote an empty box by `$0$' and a filled box by `$1$'. 

\begin{figure}[htbp]
\begin{center}
\begin{picture}(0,0)
\put(32,65){$Q_0^0$}
\put(116,65){$Q_1^0$}
\put(153,65){$Q_2^0$}
\put(75,60){$W_0^0$}
\put(135,65){$W_1^0$}
\put(180,60){$W_2^0$}
\put(70,-4){$Q_0^1$}
\put(135,-4){$Q_1^1$}
\put(170,-4){$Q_2^1$}
\put(25,0.5){$W_2^1$}
\put(113,0.5){$W_0^1$}
\put(153,-12){$W_1^1$}
\end{picture}
\includegraphics[viewport=103 585 486 710, clip,
width=6.5cm,height=2.3cm]{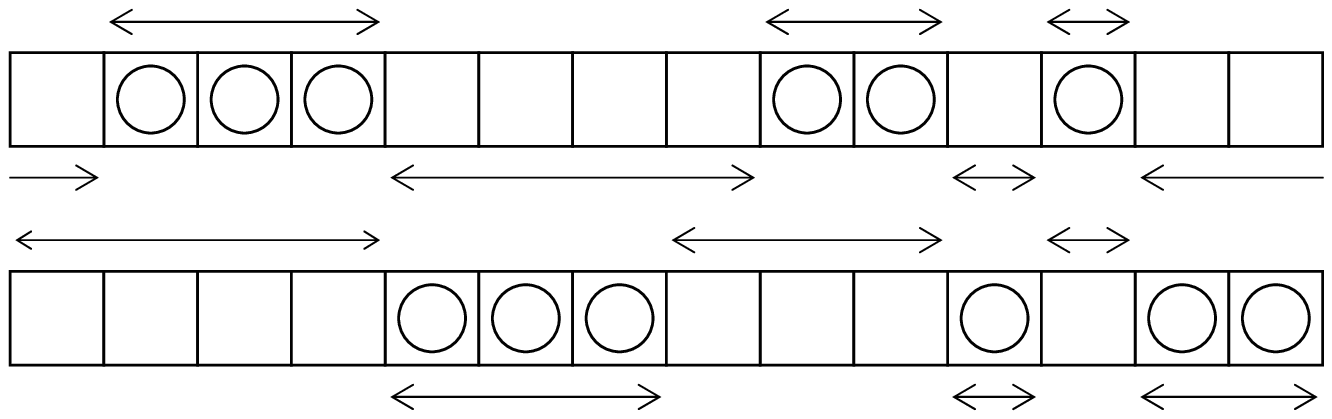}
\end{center}
\caption[ji]{The definition of $Q_j^t$ and $W_j^t$}
\label{naming2}
\end{figure}

Let $N$ be the number of groups of consecutive `$1$'s at $t=0$. (Clearly,
$N$ is also the number of groups of
$\mbox{ of consecutive `$0$'s at $t=0$}$). This number $N$ does not change
under the time evolution 
and it corresponds to the number of solutions in the pBBS.
We introduce dependent variables $Q_j^t$, $W_j^t(j=1,\dots,N,t$
$\in\NN)$, as in figure  \ref{naming2}.\\[3mm]
At $t=0$, choose one of the sets of consecutive `$1$'s and denote the number of 
`$1$'s by $Q_0^0$. Next, looking to the right, denote the number of `$0$'s 
in the nearest set of consecutive `$0$'s by $W_0^0$. Then, looking to the right, 
denote the number of `$1$'s 
in the nearest set of consecutive `$1$'s by $Q_1^0$. We continue to define 
$W_1^0,Q_2^0,\dots,Q_N^0,W_N^0$ in a similar manner. 
Since our system has the periodic boundary condition, 
it follows $Q_{N}^0=Q_0^0$,  
$W_{N}^0=W_0^0, \dots etc$. 
In the following, we always use the convention that the position $j$ is
defined in $\ZZ_N$ (i.e. $Q_{j+N}^t=Q_{j}^t$, $W_{j+N}^t=W_{j}^t$).
At $t=1$, to define $Q_0^1,W_0^1,\dots$,etc., one needs to choose
one of the sets of consecutive `$1$'s;
The set of consecutive `$1$'s whose leftmost entry was updated from the `$0$'
of the first set of consecutive `$0$'s will be called $Q_0^1$.
In general, $Q_0^{t+1}$ is defined as the number of entries in the set of
consecutive `$1$'s whose leftmost entry was updated from the `$0$' of
the first set of consecutive `$0$'s at $t$.

The following formulae describe the time evolution of the pBBS.
\begin{eqnarray}
Q_i^{t+1}&=&\min{[W_i^t,X_i^t+Q_i^t]}\label{hako}\\
W_i^{t+1}&=&Q_{i+1}^t+W_i^t-Q_i^{t+1}\label{tama} \\
X_i^t&=&\!\!\!\!
\max_{k=0,\dots,N-1}{\left[\sum_{l=1}^k{(Q_{i-l}^t-W_{i-l}^t)}\right]},\label{kei}
\end{eqnarray}
where it follows that
\begin{equation}
\sum_{i=1}^N{Q_i^t}<\sum_{i=1}^N{W_i^t}\label{tamakazu}
\end{equation}
due to the condition $Q<L/2$.

The main feature we use to solve the initial value problem of pBBS is 
the correspondence between the pBBS and the periodic Toda equation.
\begin{defi}\label{pdToda} 
The periodic Toda equation (pd Toda equation) is given as
\begin{eqnarray}
I_i^{t+1}&=&I_i^t+V_i^t-V_{i-1}^{t+1}\label{toda1}\\
V_i^{t+1}&=&\frac{I_{i+1}^tV_i^t}{I_i^{t+1}}\label{toda2}
\end{eqnarray}
with the boundary condition 
\begin{equation}
I_{i+N}^t=I_i^t,\quad\ \ V_{i+N}^t=V_i^t.\label{todaperi}
\end{equation}
\end{defi}
The following proposition shows the essential
relation between the pBBS and the pd Toda equation.
\begin{prop}[\cite{kimijima}]\label{niiteniti}
Suppose that the pd Toda equation in definition \ref{pdToda}
satisfies the condition 
\begin{equation}
0<\prod_{i=1}^N{V_i^t}<\prod_{i=1}^N{I_i^t}.\label{todakazu}
\end{equation}
If the pd Toda equation has a one parameter family
of solutions $I_j^t(\ep)$ and $V_j^t(\ep)$\,(for parameter $\ep$),
then if
\[
Q_j^t\equiv\lim_{\ep\to +0}{-\ep\log{I_j^t(\ep)}}\,\,\,\ \ \mbox{and}\,\,\,\ \
W_j^t\equiv\lim_{\ep\to +0}{-\ep\log{V_j^t(\ep)}}
\]
exist, they
satisfy the equations (\ref{hako}), (\ref{tama}), (\ref{kei}),
and (\ref{tamakazu}).
\end{prop} 
\proof Substituting (\ref{toda2}) to (\ref{toda1}), we have
\[
I_i^{t+1}=I_i^t+V_i^t-\frac{I_i^tV_{i-1}^t}{I_{i-1}^{t+1}}.
\]
Since $I_{i-1}^{t+1}$ satisfies the same equation,
\begin{eqnarray*}
I_{i}^{t+1}=I_i^t+V_i^t-\frac{I_i^tV_{i-1}^t}
{I_{i-1}^t+V_{i-1}^t-\frac{I_{i-1}^tV_{i-2}^t}{I_{i-2}^{t+1}}}
\end{eqnarray*}
Repeating this procedure, we get the following equation due to the 
periodic boundary condition. 
\begin{eqnarray*}
I_i^{t+1}= I_i^t+V_i^t-\frac{I_i^tV_{i-1}^t}
{I_{i-1}^t+V_{i-1}^t-\frac{I_{i-1}^tV_{i-2}^t}
{I_{i-2}^t+V_{i-2}^t-\frac{I_{i-2}^tV_{i-3}^t}
{I_{i-3}^t+V_{i-3}^t-\frac{I_{i-3}^tV_{i-4}^t}
{\ddots\frac{I_{i+1}^tV_{i}^t}{I_{i}^{t+1}}}}}}.
\end{eqnarray*}
This is a quadratic equation of $I_i^{t+1}$. The two solutions are
\begin{eqnarray*}
I_i^{t+1}=V_i^t,
\end{eqnarray*}
and
\begin{eqnarray}
I_i^{t+1}
&=&
I_i^t\frac{1+\frac{V_i^t}{I_i^t}+\frac{V_i^tV_{i-1}^t}{I_i^tI_{i-1}^t}
+\cdots+\frac{V_i^tV_{i-1}^t\dots V_{i+2}^t}{I_i^tI_{i-1}^t\dots I_{i+2}^t}}
{1+\frac{V_{i-1}^t}{I_{i-1}^t}+\frac{V_{i-1}^tV_{i-2}^t}{I_{i-1}^tI_{i-2}^t}
+\cdots
+\frac{V_{i-1}^tV_{i-2}^t\dots V_{i+1}^t}{I_{i-1}^tI_{i-2}^t\dots I_{i+1}^t}}\\
&=&
V_i^t+I_i^t\frac{1-\frac{V_1^tV_2^t\dots V_N^t}{I_1^tI_2^t\dots I_N^t}}
{1+\frac{V_{i-1}^t}{I_{i-1}^t}+\frac{V_{i-1}^tV_{i-2}^t}{I_{i-1}^tI_{i-2}^t}
+\cdots
+\frac{V_{i-1}^tV_{i-2}^t\dots V_{i+1}^t}{I_{i-1}^tI_{i-2}^t\dots I_{i+1}^t}}.
\label{toda4}
\end{eqnarray}
The first one does not satisfy the condition (\ref{tamakazu}). The other one
gives the time evolution for $I_i^t$.

Now, we calculate the ultradiscrete limit of (\ref{toda2}) and (\ref{toda4}).

We put
$
\displaystyle I_i^t=\exp{\left[-\frac{Q_i^t(\ep)}{\ep}\right]} 
$
, $\displaystyle V_i^t=\exp{\left[-\frac{W_i^t(\ep)}{\ep}\right]}$
and take a limit $\ep\to +0$.
By virtue of the fact that \[
\lim_{\ep\to +0}{-\ep\log{(e^{-a/\ep}+e^{-b/\ep})}}=\min{[a,b]}
\] and 
\[
0<\prod_{i=1}^N{V_i^t}<\prod_{i=1}^N{I_i^t} \quad\Rightarrow\quad
\lim_{\ep\to +0}{\ep\log
{\left[1-\frac{V_1^tV_2^t\dots V_N^t}{I_1^tI_2^t\dots I_N^t}\right]}}=0,
\]
it is a straightforward result that (\ref{toda2}) yields (\ref{tama}),
and (\ref{toda4}) yields (\ref{hako}) if $Q_i^t, W_i^t$ exist.
$\qed$

We shall use this proposition to solve the initial value
problem of the pBBS. Our strategy can be summarized as follows:
\begin{enumerate}
\def\labelenumi{(\theenumi)}
\item For given initial data $Q_j^0$, $W_j^0$ ($j=1,2,\dots,N$),
we associate initial values with the pd Toda equation as
\begin{equation}\label{asso}
I_j^0=\exp{\left[-\frac{Q_j^0}{\ep}\right]},\hspace{1cm}
V_j^0=\exp{\left[-\frac{W_j^0}{\ep} \right]}.
\end{equation}
\item Then, we solve the initial value problem of the pd Toda equation 
by the inverse scattering method. The solution $\{I_j^t(\ep),
V_j^t(\ep)\}$ depends on the parameter $\ep$.
\item In principle, by prop \ref{niiteniti}, the solution to the pBBS is obtained
as 
\begin{equation}
Q_j^t=\lim_{\ep\to +0}{-\ep\log{I_j^t(\ep)}},
\hspace{1cm}W_j^t=\lim_{\ep\to +0}{-\ep\log{V_j^t(\ep)}}.
\end{equation}
\end{enumerate}
\begin{rem}\label{pertur}
One does not always have to define $I_j^0$ and $V_j^0$ as (\ref{asso}).
Indeed, one could define these numbers freely on the condition that
\begin{equation}
Q_j^0=\lim_{\ep\to +0}{-\ep\log{I_j^0(\ep)}},
\hspace{1cm}W_j^0=\lim_{\ep\to +0}{-\ep\log{V_j^0(\ep)}}.
\end{equation}
\end{rem}
\subsection{Solution of pd Toda equation}\label{abesi2}
The initial value problem of the pd Toda equation was solved by 
the algebro-geometric method \cite{kimijima}.
In this paper, we omit most of
the details of the method and give only the solution.

\begin{figure}[htbp]
\begin{center}
\begin{picture}(0,0)
\put(120,70){$a_1$}
\put(160,70){$a_2$}
\put(200,70){$a_3$}
\put(185,45){$b_3$}
\put(135,51){$b_2$}
\put(92,55){$b_1$}
\end{picture}
\includegraphics[viewport=137 527 512 699, clip,
height=3cm,width=10cm]{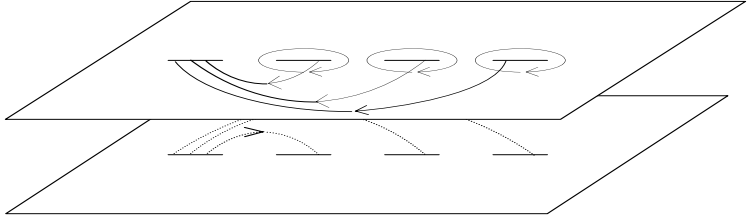}
\end{center}
\caption[Homology]{Canonical basis of $H_1(C,\ZZ)$.\ (Case for $g=3$.)}
\label{canbasis}
\end{figure}

Let $C$ be a hyperelliptic curve of genus $g$, and define the base of 
$H_1(C,\ZZ)$ as in figure \ref{canbasis}.
We denote the normalized 1-form on $C$ by $\{\omega_i\}_{i=1}^g$,
and the period matrix of $C$ by $B=\left(\int_{b_i}{\omega_j}\right)_{i,j}$.
\begin{rem}
$\{\omega_i\} \mbox{ is normalized } \,\Leftrightarrow\,
\int_{a_i}{\omega_j}=\delta_{i,j}\quad \forall i,j.$
\end{rem}
A hyperelliptic curve $C$ of degree $g$ can be exprresed as 
\[
\mu^2=G(\lam),\]
where $G(\lam)$ is a polynomial in $\lam$
of degree $2g+1$ or $2g+2$. 
Any holomorphic differential on $C$
can be rewritten as
\[
c_{g-1}\frac{\lam^{g-1}\,d\lam}{\mu}+c_{g-2}\frac{\lam^{g-2}\,d\lam}{\mu}
+\cdots+c_0\frac{d\lam}{\mu},\quad c_0,\dots,c_{g-1}\in\CC.
\]
Let us define the complex constants $c_{j,k}(j=1,2,\dots,g,\,
k=0,1,\dots,g-1)$ as
\[
\omega_j=c_{j,g-1}\frac{\lam^{g-1}\,d\lam}{\mu}
+c_{j,g-2}\frac{\lam^{g-2}\,d\lam}{\mu}
+\cdots+c_{j,0}\frac{d\lam}{\mu}.
\]

The Abelian mapping and the theta function are the most important tools 
in the method.
We define the quotient space 
 $\CC^g/(\ZZ^g\!\! +\!\! B\ZZ^g)$ obtained by the equivalence relation
\[
x\sim y \in \CC^g \Leftrightarrow x-y \in \ZZ^g\! \! + \!\! B\ZZ^g\ .
\]
For a fixed point $P_0\in C$, the mapping 
\[
C\ni P \mapsto \int_{P_0}^{P}{\vect{\omega}} \equiv 
\left(\int_{P_0}^P{\omega_1},\dots,\int_{P_0}^P{\omega_g}  \right) \in 
\CC^g/(\ZZ^g\!\! +\!\! B\ZZ^g)
\]
is a well-defined Abelian mapping.
The Abelian mapping is usually denoted by
\[
\vect{A}(P)=\int_{P_0}^{P}{\!\!\vect{\omega}}\, .
\]
The Abelian mapping of a divisor $D=\sum_i{n_iP_i}$ is defined by the formula
\[
\vect{A}(D)=\sum_i{\, n_i\!\int_{P_0}^{P_i}{\!\!\vect{\omega}}}.
\]

\begin{defi}\label{theta}
Let $B$ be a $g\times g$ matrix which satisfies the relation 
\[
B=B^t\quad\quad \mbox{and}\quad\quad \mathrm{Im}B>0.
\]
Then the theta function $\theta(\vect{z},B)$  for $\vect{z}\in\CC^g$ is 
defined as the 
holomorphic function
\[
\theta(\vect{z},B)=\sum_{\vect{n}\in\ZZ^g}
{\exp{(\pi \rmi \vect{n}^tB\vect{n}+2\pi \rmi\vect{n}^t\vect{z})}}.
\]
\end{defi}
\begin{rem}
The theta function $\theta(\vect{z},B)$ satisfies
\[
\theta(\vect{z}+\vect{e}_j,B)=\theta(\vect{z},B),\quad
\vect{e}_j=(0,0,\cdots,\mathop{\hat{1}}^{j},\cdots,0)^t,
\]
and
\[
\theta(\vect{z}+\vect{b}_j,B)=e^{-2\pi \rmi z_j-\pi \rmi B_{jj}}
\theta(\vect{z},B),\quad \vect{b}_j=(B_{1j},\cdots,B_{gj})^t.
\]
\end{rem}

In our algebro-geometric method (i.e.\relax\ inverse scattering method),
we use these functions and Abelian integrals 
on some hyperelliptic curve $C$
to describe the solution of pd Toda equaiton. 

To define the curve $C$ associated to the initial condition
$\{I_j^0,V_j^0\}_{j=0}^{N-1}$, we prepare two sequences $\{x_n\}_{n\in\ZZ}$
and $\{y_n\}_{n\in\ZZ}$ by (\ref{kisei}) and (\ref{kisei2}).
\begin{equation}\label{kisei}
\left\{
\begin{array}{c}
\!\!x_{n+1}\!=\!\{\lam-(I_{n+1}^0+V_n^0)\}x_n-(I_n^0V_n^0)x_{n-1}\\
\!\!y_{n+1}\!=\!\{\lam-(I_{n+1}^0+V_n^0)\}y_n-(I_n^0V_n^0)y_{n-1}
\end{array}
\right.,\,
\end{equation}
\begin{equation}\label{kisei2}
\left(\begin{array}{@{\,}c@{\,}}
	x_0 \\
	x_1
\end{array}\right)\!=\!
\left(\begin{array}{@{\,}c@{\,}}
	0 \\
	1
\end{array}\right),\,
\left(\begin{array}{@{\,}c@{\,}}
	y_0 \\
	y_1
\end{array}\right)\!=\!
\left(\begin{array}{@{\,}c@{\,}}
	1 \\
	0
\end{array}\right).
\end{equation}

Let $C$ be a hyperelliptic curve defined by
\begin{equation}\label{hirosima}
\mu^2=\Delta(\lam)^2-4m^2,
\end{equation}
where 
\begin{equation}\label{tokyo}
\Delta(\lam)=x_{N+1}+y_N,\quad m^2=\prod_{i=1}^N{I_i^0V_i^0}.
\end{equation}
Note that $\Delta(\lam)$ is a monic polynomial in $\lam$ of degree $N$.
The genus of the hyperelliptic curve $C$ is $N-1\,(=:g)$. 

Seperately form the definition of $C$, we define $N-1$ complex numbers
$\mu_j$, $(j=1,2,\dots,N-1)$ as the roots of 
\begin{equation}\label{nagasaki}
y_{N+1}=0.
\end{equation}
Note that $y_{N+1}$ is a polynomial of degree $N-1$, the highest coefficient of which
is $-I_1^0V_1^0$.
\begin{thm}[\cite{kimijima}]
Let $C$ be a hyperelliptic curve (\ref{hirosima}),
and define the canonical basis of $H_1(C,\ZZ)$ as in figure \ref{canbasis}.
Then,
the solution of
the pd Toda equation (\ref{toda1}) \& (\ref{toda2}) is expressed as follows.
\begin{eqnarray}
I_{n+2}^t+V_{n+1}^t\!\!&=&\!\!
\sum_{j=0}^{g}{\lam_j}-\sum_{j=1}^g{\int_{a_j}
{\!\lam\,\omega_j}}\nonumber\\
&&\hspace{9mm}
-\sum_{j=1}^g{c_{j,g-1}\frac{d}{du_j}\log{\frac{\theta(n\vect{r}
+t\vect{\nu}+\vect{c})}{\theta((n+1)\vect{r}+t\vect{\nu}+\vect{c})}
}}\label{solutionofToda},
\end{eqnarray}
where
$
\vect{r}=\vect{A}(\infty^--\infty^+)$,
$
\vect{\nu}
=\vect{A}(0-\infty^+),
$
$\displaystyle\vect{c}(0)=\vect{A}(\infty_+-\sum_{j=1}^g{P_j^0})-\vect{K}$.\\
$\infty^+$ is the point at
infinity in the upper sheat of $C$, and $\infty^-$ is
the point at infinity in the lower sheat. $0$ is the point in the lower sheat with
$\lam(0)=0$. $\vect{K}$ is a Riemann constant of $C$ (\cite{Munford},\cite{Fay}),
and $\lam_0,\lam_1,\dots,\lam_g$ are the roots of $\Delta(\lam)=0$.
And $P_j$ is a point on $C$, which satisfies $\lam(P_j)=\mu_j$.
\end{thm}
\subsection{Young diagram}\label{paiotu} 
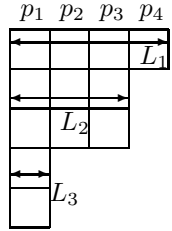
\begin{figure}[htbp]
\begin{center}
\begin{picture}(100,90)(0,0)
\put(4,80){$p_1$}
\put(19,80){$p_2$}
\put(34,80){$p_3$}
\put(49,80){$p_4$}
\put(49,62){$L_1$}
\put(19,37){$L_2$}
\put(15,10){$L_3$}
\put(0,0){\line(0,1){75}}
\put(0,0){\line(1,0){15}}
\put(15,0){\line(0,1){75}}
\put(0,15){\line(1,0){15}}
\put(0,30){\line(1,0){45}}
\put(30,30){\line(0,1){45}}
\put(45,30){\line(0,1){45}}
\put(0,45){\line(1,0){45}}
\put(0,60){\line(1,0){60}}
\put(0,75){\line(1,0){60}}
\put(60,60){\line(0,1){15}}
\put(0,20){\vector(1,0){15}}
\put(15,20){\vector(-1,0){15}}
\put(0,49){\vector(1,0){45}}
\put(15,49){\vector(-1,0){15}}
\put(0,70){\vector(1,0){60}}
\put(15,70){\vector(-1,0){15}}
\end{picture}
\end{center}
\caption{The Young diagram associated to the state in figure \ref{0-solitons}}
\label{Young tab}
\end{figure}

In this section, we briefly review the correspondence
between box-ball systems and Young diagrams.
A Young diagram is a collection of boxes as shown in figure \ref{Young tab}.
We define a Young diagram associated to a state of pBBS.

\begin{figure}[htbp]
\begin{picture}(100,10)(0,0)
\thicklines
\put(106,-10){\line(0,1){4}}
\put(106,-6){\line(1,0){8}}
\put(114,-10){\line(0,1){4}}
\put(156,-10){\line(0,1){4}}
\put(156,-6){\line(1,0){8}}
\put(164,-10){\line(0,1){4}}
\put(181,-10){\line(0,1){4}}
\put(181,-6){\line(1,0){8}}
\put(189,-10){\line(0,1){4}}
\put(197,-10){\line(0,1){4}}
\put(197,-6){\line(1,0){8}}
\put(205,-10){\line(0,1){4}}
\put(248,-10){\line(0,1){4}}
\put(248,-6){\line(1,0){8}}
\put(256,-10){\line(0,1){4}}
\thinlines
\put(97,-10){\line(0,1){8}}
\put(97,-2){\line(1,0){25}}
\put(122,-10){\line(0,1){8}}
\put(239,-10){\line(0,1){8}}
\put(239,-2){\line(1,0){25}}
\put(264,-10){\line(0,1){8}}
\put(173,-10){\line(0,1){8}}
\put(173,-2){\line(1,0){41}}
\put(214,-10){\line(0,1){8}}
\put(89,-10){\line(0,1){12}}
\put(89,2){\line(1,0){42}}
\put(131,-10){\line(0,1){12}}
\put(148,-10){\line(0,1){12}}
\put(148,2){\line(1,0){75}}
\put(223,-10){\line(0,1){12}}
\put(231,-10){\line(0,1){12}}
\put(231,2){\line(1,0){42}}
\put(273,-10){\line(0,1){12}}
\put(81,-10){\line(0,1){16}}
\put(81,6){\line(1,0){58}}
\put(139,-10){\line(0,1){16}}
\end{picture}
\begin{center}
0\ 0\ 0\ 0\ 0\ 1\ 1\ 1\ 1\ 0\ 0\ 0\ 0\ 1\ 1\ 0\ 1\ 1\ 0\ 1\ 0\ 0\relax
\ 0\ 1\ 1\ 1\ 0\ 0\ 0\ 0\ 0\ 0\ 0
\end{center}
\begin{center}
$\Downarrow$
\end{center}
\begin{center}
0\ 0\ 0\ 0\ 0\ 1\ 1\ 1\ 0\ 0\ 0\ 1$\vert$1$\vert$\relax
0\ 0\ 1\ 1\ 0\ 0\ 0\ 0\ 0\ 0
\end{center}
\caption[10elimination]
{An  example of 10-elimination. A $3$-soliton system with two 0-solitons
is obtained from a $5$-soliton system.}
\label{0-solitons}
\end{figure}
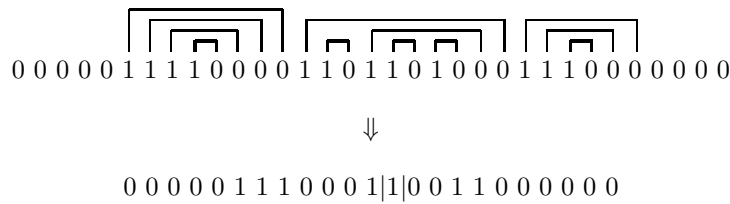

Let us consider the pBBS which has $N$-solitons (Section \ref{pBBS}).
When we regard
the pBBS as a dynamical system of a finite sequence of
`$0$'s and `$1$'s, we can introduce the following operation 
which we shall call `10-elimination'.
\begin{enumerate}
\def\labelenumi{(\theenumi)}
\item For a given state, connect all the `10' pairs in the sequence with arc lines.
\item Neglecting the 10 pairs which were connected in the first step,
connect all the remaining `10's with arc lines.
\item Repeat the above procedure until all the `1's are connected to `0's.
\item Eliminate the `10's in a state, and obtain a new sequence.
\end{enumerate}
Figure \ref{0-solitons} shows an example of 10-elimination.
 The mark `$\vert$' means `0-soliton' which has no entry but has a position.
A 0-soliton appears when we eliminate a soliton of length $1$.
We can perform
this `10-elimination' repeatedly and transform any $N$-soliton system
into a $(N-k)$-soliton system with $k$ 0-solitons.
Note that $k$ is the number of the shortest solitons in the $N$-soliton 
state. Note also that the 0-solitons do not move under the time evolution rule.

Let $p_1$ be the number of 10 pairs in a state of the pBBS, connected with arc lines 
in the first step of 10-elimination, (i.e.\relax\ after one elimination).
Similarly, we denote by $p_j$ the number of 10
pairs in the $j$-th step of 10-elimination. Note that $p_1\geq p_2\geq \dots\geq p_l$,
where $l$ is the number of the last step. The most important aspect of
these integers $p_j$ is the fact that the series $\{p_1,p_2,\dots,p_l\}$ are
conserved quantities for the time evolution of the pBBS \cite{Yoshihara}.
Using this series,
we can associate a state of the pBBS with a Young diagram
with $p_j$ boxes in the $j$-th column $(j=1,2,\dots,l)$ (see figure \ref{Young tab}). 
Then let us denote the distinct lengths of the rows by
$\{L_1,L_2,\dots,L_s\}$. Note that $L_1>L_2>\dots>L_s$.

The following is a main theorem in this paper.
\begin{thm}\label{renkon}
Let $C : \mu^2=\Delta(\lam)^2-4m^2$ be the hyperelliptic curve 
associated with the initial value problem of pd Toda equation defined by
(\ref{toda1}),(\ref{toda2}),(\ref{todaperi}),(\ref{todakazu}), and let
\[
I_n^0:=e^{-\frac{Q_n^0}{\ep}}\quad\mbox{and}\quad V_n^0:=e^{-\frac{W_n^0}{\ep}}
\]
(see section \ref{abesi2}).
Then all of the roots of $\Delta(\lam)^2-4m^2=0$ are simple and positive.

Let us denote these by
\[
0<\lam_0^-<\lam_0^+<\lam_1^-<\lam_1^+<\dots<\lam_g^-<\lam_g^+,\ (g=N-1).
\]
Moreover,
\[\; 
-\lim_{\ep\to +0}{\ep\log{\lam_j^\pm}}=
\left\{
\begin{array}{l}
\mbox{the length of the $(N-j)$-th row of the}\\
\mbox{Young diagram associated with the state } \{Q_n^0,W_n^0\}
\end{array}
\right\}.
\]
\end{thm}

\begin{figure}[htbp]
\begin{center}
\begin{picture}(322,100)(0,10)
\put(0,70){\framebox(90,35)
{$\begin{array}{c}
\mbox{A state of pBBS} \\
\{Q_n^0,W_n^0\}
\end{array}$}}
\put(90,80){\vector(1,0){100}}
\put(190,70){\framebox(170,35)
{$\begin{array}{c}
\mbox{pd Toda equation (\ref{toda1})-(\ref{todakazu})} \\
\mbox{with initial condition }\{I_n^0,V_n^0\}
\end{array}$}}
\put(93,95)
{$
\begin{array}{c}
I_n^0=\exp{-Q_n^0/\ep},\\
V_n^0=\exp{-W_n^0/\ep}
\end{array}
$}
\put(35,70){\vector(0,-1){35}}
\put(270,70){\vector(0,-1){35}}
\put(40,50){Conserved quantity}
\put(170,50){Conserved quantity}
\put(0,15){\framebox(90,20)
{Young diagram}}
\put(190,15){\framebox(170,20)
{Curve 
$C : \mu^2=\Delta(\lam)^2-4m^2$}}
\put(190,23){\vector(-1,0){100}}
\put(93,26){Ultradiscretization}
\put(104,10){\underline{Theorem \ref{renkon}}}
\end{picture}
\end{center}
\caption{The relation between the conserved quantities of two systems,
the pBBS and the pd Toda equation.}
\end{figure}
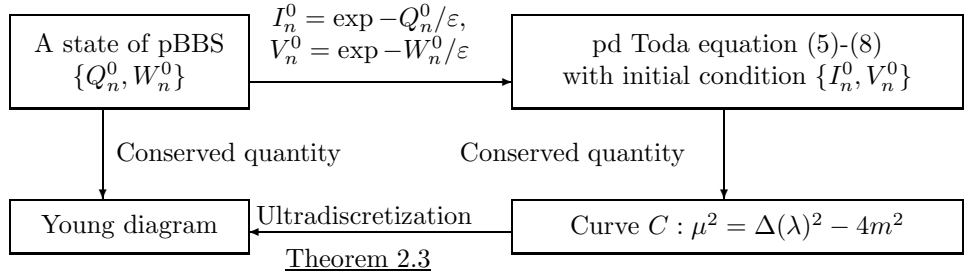

By virture of theorem \ref{renkon}, the ultradiscrete limits of the data which are described
by $\lam_j^\pm\, (j=0,1,\dots,g)$ can be expressed in terms of the Young diagram. 
In fact, several essential data for the pd Toda equation
---fundamental period etc.--- can be expressed
by $\lam_j^\pm$s only. Hence, we can express the ultradiscrete limit of these data
by the information of the associated Young diagram.
\section{Proof of theorem \ref{renkon}}\label{ultra} 
\subsection{Ultradiscrete limit}
For convenience, we first state several lemmas which we will
use in the rest of the present paper.
\begin{defi}
Let $\varepsilon$ be a positive number, and let $f(\varepsilon)$
be a continuous real-valued function of $\varepsilon$. Let us introduce the symble
`$\ud$' which means
\[
f \ud F \ \Leftrightarrow\ 
\left\{
\begin{array}{ll}
\exists \delta>0 \ \ \ \mbox{s.t.}\ \ \  0<\ep<\delta \ \Rightarrow\   f(\ep)>0\\
 F=\lim_{\ep\to 0}\ep\log{f(\ep)}
\end{array} .
\right.
\]
\end{defi}
\begin{rem}\label{oh}
By definition, $f\ud F \ \Leftrightarrow\  \log{f}\sim F/\ep \ \ (\ep\to 0)$.
\end{rem}
\begin{defi}\label{defgo}
The sign `$\udsim$$\!$' stands for the relation
\[
f \udsim g \ \Leftrightarrow \ f \ud F ,\  g\ud G\    \mbox{and} \ F=G.
\]
\end{defi}
\begin{rem}\label{relation of ud and sim}
The sign `$\udsim$$\!$' is not equivalent with the usual sign $`\sim\!\!$ '\, 
\mbox{$\mathrm{i.e.}$}\, ${f\sim g}\ \Leftrightarrow\ 
\exists m,M>0\ \mbox{s.t.}\ 
0\! < \! m \!
<\! \lim_{\ep\to 0}{\vert f(\ep)/g(\ep)\vert}\! <\! M \! <\! +\infty. $ For 
example, if 
$f(\ep)=\ep^{-1},\, g(\ep)=\ep^{-2},$\, then $f \udsim g$ but $f \nsim g$.
Precisely speaking, under the condition $f(\ep) , \, g(\ep)\! >\!0$
for $0<\!\ep\!<\!\!\!<1$,
$f\! \sim\! g  \Rightarrow  f\udsim g$ holds, but the
inverse relation does not necessarily
hold.
\end{rem}
\begin{lemma}\label{wa}
Let $f_1,f_2,\dots,f_N,g$ be continuous real-valued functions of $\ep$ with 
$g=f_1+f_2+\cdots+f_N$. If \[
g\ud G,\ f_1\ud F_1,\ \dots,\ f_N\ud F_N,
\]
then $G=\max{\{F_1,\dots,F_N  \}}$.
\end{lemma}
\proof By definition, for some number $\delta>0$, 
$f_1(\ep)\, ,\dots,\, f_N(\ep)\, ,g(\ep)$ are all positive if $\ep\in (0,\delta)$.
It can be assumed that $F_1\geq F_2\geq \dots \geq F_N$ without 
loss of generality. For the largest numbers $F_1=F_2=\dots=F_m\ (1\leq m\leq N)$,
one can rearrange the index if necessary and assume \[
f_1\succeq f_2 \succeq \dots\succeq f_m,
\]
where $h\succeq k \stackrel{\mathrm{Def}}{\Longleftrightarrow} 
\exists C>0,\,\lim_{\ep\to 0}{\zet{k(\ep)/h(\ep)}}<C<+\infty $.
So, $g\sim f_1$ is obvious, therefore, 
noticing remark \ref{relation of ud and sim} and Def.\ref{defgo}, 
$G=F_1$ is proved.$\qed$

We have the following obvious lemma
\begin{lemma} \label{seki}
Let $f,g$ be continuous real-valued functions of $\ep$.
\[
f\ud F,\ g\ud G \ \Rightarrow\ fg\ud F+G. 
\]
\end{lemma}

\begin{lemma}\label{asym}
Let $f,g,h$ be continuous real-valued functions of $\ep$
with $f(\ep)\!\!=\!\!g(\ep)+h(\ep)$.
If $f \ud F,\ \zet{g} \ud G, 
\, \zet{h} \ud H$, and $G \neq H$, then $F=\max{[G,H]}$.
\end{lemma}
\proof Without loss of generality, one can assume $G>H.$ Let $\delta$ be 
a positive number which admits 
\[
0<\ep <\delta \,\Rightarrow \, 
\left\{
\begin{array}{ll}
f(\ep)>0\  \\
   \zet{h(\ep)}
<C_\delta \zet{g(\ep)}
\end{array}\\
\right.
\]
$\mbox{for some}\ C_\delta \ 
\mbox{which depends
 on only}\ \delta\ 
\mbox{and $C_\delta\to 0\ (\delta\to 0)$.}$

Thus, $0< f\leq \zet{g}+\zet{h}<(1+C_\delta)\zet{g}
\ \ \mbox{for}\ \ \ep\in (0,\delta)$, and $F\leq G$ holds. 
On the other hand, the
inequality $0<\zet{g}\leq \zet{f}+\zet{h}$ gives 
\[
1\leq \frac{\zet{f}}{\zet{g}}+\frac{\zet{h}}{\zet{g}} <
\frac{\zet{f}}{\zet{g}}+C_\delta\ \ \mbox{for}\ \ \ep\in (0,\delta).
\]
As $C_\delta\to 0$ for decreasing $\delta$,
it follows that $F\geq G$.$\qed$
\begin{rem}\label{mahlar}
If we omit the condition `$G\neq H$', the claim of lemma \ref{asym}
becomes
\[
F\leq \max{[G,H]}.
\]
\end{rem}

\begin{lemma}\label{poly}
Let $f(\lambda,\ep)$ be a polynomial in $\lam$ with 
real coefficients of degree $N+1$:
\[
f(\lambda,\ep)=\lambda^{N+1}-k_N(\ep)\lam+k_{N-1}(\ep)\lam^{N-1}-\dots+
(-1)^{N+1}k_0(\ep)
\]
where $k_j(\ep)>0 \ (j=0,1,\dots,N)$ and $k_j \ud K_j$. 
Then, the roots of the equation 
$f(\lam,\ep)=0$, $\lam_0(\ep)<\lam_1(\ep)<\dots <\lam_g(\ep)$,
satisfy 
\[
\lam_N\ud K_N,\  \lam_{N-1}\ud K_{N-1}-K_{N},\ 
\dots,\lam_0\ud K_{0}-K_1.
\]
\end{lemma}
\proof The fundamental relation between roots and coefficients gives 
\begin{eqnarray*}
&&k_N\!=\!\lam_0+\lam_1+\cdots+\lam_N ,\\
&& k_{N-1}\!=\! \lam_0\lam_1\!+\cdots+\lam_{N-1}\lam_N, \\
&&\dots,\\  
&&k_0=\lam_0\lam_1\dots \lam_N\!.
\end{eqnarray*}
Using lemma \ref{wa} and \ref{seki}, it is easy to prove the lemma.$\qed$
\subsection{Ultradiscretization of the polynomial $\Delta(\lam)$}
In this subsection, 
we define and calculate the key parameters associated to an initial state of the pBBS
denoted by $U_j,\,(j=0,1,\dots,N-1)$ and $P_k,\,(k=0,1,\dots,N-2)$.
In subsequent subsections, the ultradiscrete limit of the solution of 
pd Toda equation (\ref{solutionofToda}) is expressed by $U_j$ and $P_k$.

Let $C:\mu^2=\Delta(\lam)^2-4m^2$ be the hyperelliptic curve defined by (\ref{hirosima}).
Note that $\Delta(\lam)$ is a monic polynomial in $\lam$ of degree $N(=g+1)$.

We use the following two propositions without the proof.

\begin{prop}
The roots of
\[
\Delta(\lam)=0
\]
are all real and positive.
\end{prop}
\proof The proof is given in \cite{Toda}. $\qed$ 
\begin{prop}
The $2N$ roots of the equation 
\begin{equation}\label{iti}
\Delta(\lam)^2-4m^2=0
\end{equation}
 are real and positive. All of these roots are simple roots.
\end{prop}
\proof The proof is given in \cite{Toda}.
$\qed$

\begin{defi}\label{definitionofU}
Let us denote $\Delta(\lam)$ by
\[
\Delta(\lam)=\lam^{g+1}-u_g\lam^g+u_{g-1}\lam^{g-1}-\cdots+(-1)^{g+1}u_0.
\]
We define the real numbers $U_j\,(j=0,1,\dots,g)$, as
\[
U_j:=-\lim_{\ep\to+0}\ep\log{u_j},
\]
or equivalently, $u_j\ud -U_j$.
\end{defi}

To define $P_k$, let us consider the polynomial $y_{N+1}(\lam)$ ((\ref{nagasaki})).
Note that $y_{N+1}(\lam)$ is a polynomial of degree $N-1$ and
the roots of $y_{N+1}(\lam)=0$ are $\mu_k\,(k=1,2,\dots,g)$.
\begin{defi}\label{definitionofP}
Let us denote $y_{N+1}(\lam)$ as
\[
y_{N+1}=-I_1^0V_1^0
\left\{\lam^{g}-v_{g-1}\lam^{g-1}+v_{g-2}\lam^{g-2}+\cdots
+(-1)^{g}v_0\hasira\right\}
\]
(see (\ref{nagasaki})).
We define the real numbers $P_k\,(k=0,1,\dots,g-1)$, as
\[
P_k:=-\lim_{\ep\to+0}\ep\log{v_k},
\]
or equivalently, $v_k\ud -P_k$.

\end{defi}

To calculate $U_j$, we need to prepare a few 
notations. 

Let the set $\mathcal{A}(N)$ be
\begin{eqnarray*}
\; \mathcal{A}(N)\!:=\!\!\left\{\!\!\!\!
\begin{array}{r}
\left\{a_1-1,a_1,\dots,a_k-1,a_k\right\}\\
\in 2^{\ZZ/ N\ZZ}
\end{array}
\!\!\right.
\left\vert
\begin{array}{cc}
k=1,2,\dots,[\frac{N}{2}]\\
\forall i,j,\ a_i \not\equiv a_j-1 \,\land\,
i\neq j\Rightarrow a_i\not\equiv a_j
\end{array}
\!\!\!\right\}
\cup \{\emptyset\}.
\end{eqnarray*}
An element of $\mathcal{A}(N)$ is a subset of $\ZZ/N\ZZ$, which consists 
of pairs of two consecutive numbers. (In $\ZZ/N\ZZ$, 
we regard $N$ and $1$ are consecutive numbers.)
For $N\geq 3$,
the number of elements of $\mathcal{A}(N)$ is equal to $F_N+F_{N-2}$,
where $F_N$ is the $N$-th Fibonacci number.
($F_{N+2}=F_{N+1}+F_N$,\ $F_1=1,F_2=2$).
\begin{prop}\label{elem}
The polynomial $\Delta(\lam)=x_{N+1}+y_N$ is expressed as
\[
\Delta(\lam)=\sum_{(j_1-1,j_1,\dots,j_k-1,j_k)\in\mathcal{A}(N)}
{Y_{j_1}\dots Y_{j_k}X_{i_1}\dots X_{i_{N-2k}}}
\]
where $\{j_1-1,j_1,\dots,j_k-1,j_k\}\sqcup \{i_1,\dots,i_{N-2k}\}=\ZZ/N\ZZ $,
and $Y_j=-I_jV_j$, $X_i=\lam-(I_{i+1}+V_i)$.
\end{prop}
The proof of proposition \ref{elem} is elementary though slightly involved.
We therefore defer the proof to the Appendix.
Defining $a_i:=I_{i+1}+V_i$ and $b_i:=I_iV_i$,\, we find:
\begin{prop}\label{zero}
The coefficients of the polynomial
\[\Delta(\lam)=\lambda^{g+1}-u_{g}\lam^{g}+u_{g-1}\lam^{g-1}-\cdots+
(-1)^{g+1}u_0\]
satisfy
\begin{equation}
u_0\ud -(Q_1+Q_2+\cdots+Q_N)
\end{equation}
and
\begin{equation}
u_g\ud -\min{[Q_i,W_i]}.
\end{equation}
Equivalently, $U_0=Q_1+Q_2+\cdots+Q_N$ and $U_g=\min{[Q_i,W_i]}$.
\end{prop}
\proof It is sufficient to prove 
\begin{eqnarray}
u_0&=&
\sum_{(j_1-1,j_1,\dots,j_k-1,j_k)\in\mathcal{A}(N)}
{\!\!\!\!(-b_{j_1})\dots (-b_{j_k})a_{i_1}\dots a_{i_{N-2k}}}\label{sumsum} \\
&=& I_1I_2\dots I_N+V_1V_2\dots V_N \label{fir}
\end{eqnarray}
and
\begin{equation}
u_g=I_1+I_2+\cdots+I_N+V_1+V_2+\cdots+V_N.\label{seco}
\end{equation}
(\ref{seco}) is a direct consequence of proposition \ref{elem}. It remains to prove
(\ref{fir}). Substituting $a_i=I_{i+1}+V_i$ and $b_i=I_iV_i$ to (\ref{sumsum}),
two types of terms appear, namely those that contain
$V_kI_k$ for some $k$ (type (i)),
and those do not (type (ii)).

Among all the terms in (\ref{sumsum}),
a contribution `$V_kI_k$' must come from 
the term which contains $-b_k$ or $a_{k-1}a_k$ only.   
For any term which contains $-b_k$, there exists one term 
in which $-b_k$ is replaced by $a_{k-1}a_k$ in (\ref{sumsum}).
Hence we can conclude that the summation of 
all terms of type (i) will cancel out.
The only terms of type (ii) are $I_1I_2\dots I_N$ and
$V_1V_2\dots V_N $ because the term of this type must come from 
$a_1a_2\dots a_N$.$\qed$

The ultradiscrete limit of $u_1,u_2,\dots,u_{g-1}$ are also 
obtained in a similar manner.
Let 
\[X=\{A_i\,\vert\, A_{2l-1}=Q_l,\, A_{2l}=W_l,\ (l=1,2,\dots,N) \}
\]
and
\[\; 
\mathcal{B}(k,N)=\left\{ 
\begin{array}{c}
\phantom{{}}\\
\!\!\!
\{A_{\sigma(i)}\}\subset X\\
\phantom{{}}
\end{array}
\right\vert\,\left. 
\begin{array}{c}
1\leq \sigma(1)<\sigma(2)<\dots<\sigma(N-k)\leq 2N \\
\sigma(i)+1<\sigma(i+1),\,\ \forall i \\
\sigma(1)=1\Rightarrow \sigma(N-k)\neq 2N
\end{array}
\right\}.
\]
\begin{prop}\label{naka}
It follows that
\[
u_k\ud -\min{\left\{
\sum_{\{A_{\sigma(i)}\}\in\mathcal{B}(k,N)}{\!\!\!\!\!\!\!\!\!\! A_{\sigma(i)}}
\,\,\right\}}\,(\equiv -U_k).
\]
\end{prop}
\proof From the proof of proposition \ref{zero}, $u_k$ can be obtained in the 
following way. First, calculate
\begin{equation}\label{yaya}
\sum_{\{i_1,i_2,\dots,i_{N-k}\}\subset \ZZ/N\ZZ}{a_{i_1}a_{i_2}\dots a_{i_{N-k}}}.
\end{equation}
And pick up the terms which contain no $V_lI_l$s. Since 
$a_l=I_{l+1}+V_l$,
a term that contains $V_lI_{l+1}$ cannot exist in (\ref{yaya}). 
Conversely, a term of length $N-k$
which has neither $V_lI_l$ nor $V_lI_{l+1}$ 
necessarily appears in (\ref{yaya}).$\qed$

We can calculate $P_0,P_1,\dots,P_{g-1}$ analogously.
\begin{prop}\label{elem2}
The polynomial $y_{N+1}(\lam)$ is of the form
\[
y_{N+1}(\lam)=-b_1\times
\!\!\!\!\sum_{(j_1-1,j_1,\dots,j_k-1,j_k)\in\mathcal{A}'(N)}
{Y_{j_1}\dots Y_{j_k}X_{i_1}\dots X_{i_{N-1-2k}}},
\]
where $X$ and $Y$ are given in proposition \ref{elem}, and
$\mathcal{A}'(N)=\mathcal{A}(N)\,\cap 2^{(\ZZ/N\ZZ-\{1\})}$
is a subset of $\mathcal{A}(N)$ of which element $(j_1-1,j_1,\dots,j_k-1,j_k)$
does not contain the number `$\,1$' $\in\ZZ/N\ZZ$.
\end{prop}
We will prove this proposition in the Appendix.

In the same way as in proposition \ref{zero} and proposition \ref{naka} we obtain:
\begin{prop}\label{xmas2}
It follows that
\[v_k\ud -\min{\left\{
\sum_{\{A_{\sigma(i)}\}\in\mathcal{B}'(k,N)}{\!\!\!\!\!A_{\sigma(i)}}
\right\}}\,(\equiv -P_k),
\]
where $\mathcal{B}'(k,N)=\mathcal{B}(k,N)\cap 2^{(X-\{Q_2,W_1\})}$ is a 
subset of $\mathcal{B}(k,N)$ which does not 
contain $Q_2$ nor $W_1$. 
\end{prop}
\proof As in the proof of proposition \ref{naka}, $v_k$ can be expressed as follows.
\[v_k=
\left[\sum_{\{i_1,i_2,\dots,i_{N-1-k}\}\subset(\ZZ/N\ZZ-\{1\})}
{\!\!\!\!\!\!\!\!\!\!a_{i_1}a_{i_2}\dots
a_{i_{N-1-k}}}\right]_{V_lI_l\to 0}
\]
Since $a_1=I_2+V_1$, we obtain the proposition.
$\qed$

The following lemma can be obtained from proposition \ref{naka} by combinatorial
arguments, which we will give in the Appendix.
Let $\bbs{N}$ be a set of an $N$-soliton state of pBBS.
\begin{lemma}\label{box-Young}
Let $x\in\bbs{N}$ be an $N$-soliton state, 
and let $U_0,U_1,\dots,U_g$ be the positive integers defined as in Def.\ref{definitionofU}.
Then $U_k$ is equal to the number of boxes 
below the $k$-th row in the Young diagram. 
\end{lemma}
\begin{figure}[htbp]
\begin{center}
\begin{picture}(145,90)(0,0)
\put(145,50){$U_0\!\!=\!\!12$}
\put(113,35){$U_1\!\!=\!\!8$}
\put(84,23){$U_2\!\!=\!\!5$}
\put(54,13){$U_3\!\!=\!\!2$}
\put(22,3){$U_4\!\!=\!\!1$}
\put(140,75){\line(1,0){3}}
\put(143,75){\line(0,-1){75}}
\put(143,0){\line(-1,0){3}}
\put(109,58){\line(1,0){3}}
\put(112,58){\line(0,-1){58}}
\put(112,0){\line(-1,0){3}}
\put(80,45){\line(1,0){3}}
\put(83,45){\line(0,-1){45}}
\put(83,0){\line(-1,0){3}}
\put(47,28){\line(1,0){3}}
\put(50,28){\line(0,-1){28}}
\put(50,0){\line(-1,0){3}}
\put(18,15){\line(1,0){3}}
\put(21,15){\line(0,-1){15}}
\put(21,0){\line(-1,0){3}}
\put(0,0){\line(0,1){75}}
\put(0,0){\line(1,0){15}}
\put(15,0){\line(0,1){75}}
\put(0,15){\line(1,0){15}}
\put(0,30){\line(1,0){45}}
\put(30,30){\line(0,1){45}}
\put(45,30){\line(0,1){45}}
\put(0,45){\line(1,0){45}}
\put(0,60){\line(1,0){60}}
\put(0,75){\line(1,0){60}}
\put(60,60){\line(0,1){15}}
\end{picture}
\end{center}
\caption{Example of the interpretation of the $U_k's$ of lemma \ref{box-Young}. }
\end{figure}
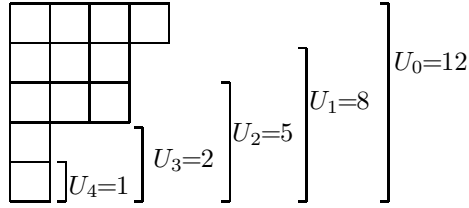

\begin{cor}
Let $x\in\bbs{N}$ be an $N$-soliton state. We can obtain the integers 
$P_0,P_1,\dots,P_{g-1}$ in Def.\ref{definitionofP} by the following procedure.
\begin{enumerate}
\def\labelenumi{(\theenumi)} %
\item Let $Q_j^0(x),W_j^0(x)\,(j=0,1,\dots,g)$ be the integers defined in section \ref{pBBS}
for a state $x\in\bbs{N}$.
Consider another $N$-soliton state $y\in\bbs{N}$, with
\[
\begin{array}{lcl}
Q_j^0(y)=Q_j^0(x) &\mbox{for}& j\neq 2, \\
W_i^0(y)=W_i^0(x) &\mbox{for}& i\neq 1, \\
\end{array}
\]
\[
Q_2^0(y)>\!\!>1,\quad\mbox{ and }\quad W_1^0(y)>\!\!>1.
\]
\item Let $p_l$ be the length of $l$-th row of the Young diagram corresponding to
$y\in\bbs{N}$.
Then,
\[
P_j=\sum_{l=j+2}^{g-1}{p_l}.
\]
\end{enumerate}
\end{cor}
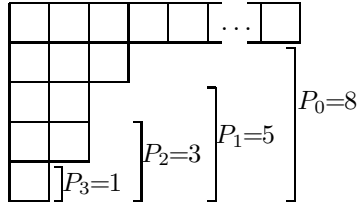
\begin{figure}[htbp]
\begin{center}
\begin{picture}(120,90)(0,0)
\put(0,0){\line(0,1){75}}
\put(0,0){\line(1,0){15}}
\put(15,0){\line(0,1){75}}
\put(0,15){\line(1,0){30}}
\put(0,30){\line(1,0){30}}
\put(30,15){\line(0,1){60}}
\put(45,45){\line(0,1){30}}
\put(0,45){\line(1,0){45}}
\put(75,60){\line(0,1){15}}
\put(0,60){\line(1,0){80}}
\put(0,75){\line(1,0){80}}
\put(60,60){\line(0,1){15}}
\put(90,60){\line(1,0){20}}
\put(90,75){\line(1,0){20}}
\put(110,60){\line(0,1){15}}
\put(95,60){\line(0,1){15}}
\put(80,65){$\dots$}
\put(105,58){\line(1,0){3}}
\put(108,58){\line(0,-1){58}}
\put(108,0){\line(-1,0){3}}
\put(75,43){\line(1,0){3}}
\put(78,43){\line(0,-1){43}}
\put(78,0){\line(-1,0){3}}
\put(47,30){\line(1,0){3}}
\put(50,30){\line(0,-1){30}}
\put(50,0){\line(-1,0){3}}
\put(17,13){\line(1,0){3}}
\put(20,13){\line(0,-1){13}}
\put(20,0){\line(-1,0){3}}
\put(109,35){$P_0\!\!=\!\!8$}
\put(78,22){$P_1\!\!=\!\!5$}
\put(50,15){$P_2\!\!=\!\!3$}
\put(20,4){$P_3\!\!=\!\!1$}
\end{picture}
\end{center}
\caption{Example of a Young diagram corresponding to a state $y$.}
\label{fololo}
\end{figure}
\proof From proposition \ref{naka} and \ref{xmas2}, it is obvious that
\[\lim_{I_2,V_1\to+0}u_{j+1}=v_j.\]
Then,
\[
P_j=-\lim_{\ep\to+0}{\ep\log{v_j}}=-\lim_{\ep\to+0}{\lim_{I_2,V_1\to+0}u_{j+1}}
=\lim_{I_2,V_1\to+0}{U_{j+1}}.
\]
The fact $I_2,V_1\to+0\,\Leftrightarrow\, Q_2,W_1\to\infty$ completes the proof.
$\qed$

\subsection{Ultradiscretization of the curve $\mu^2=\Delta(\lam)^2-4m^2$}\label{horn}
To complete the proof of theorem \ref{renkon}, we 
consider the asymptotic
behaviour of the Riemann surface $C : \mu^2=\Delta(\lam)^2-4m^2$ when $\ep\to 0$.
From (\ref{tokyo}), we easily obtain that 
$
m^2=e^{-\frac{L}{\ep}},
$
or equivalently
\begin{equation}\label{emu}
m\ud -L/2,
\end{equation}
where $L$ is the number of boxes in the pBBS (see (\ref{tokyo})).

Recall that we have denoted the roots of $\Delta(\lam)^2-4m^2=0$ by 
\[
0<\lam_0^-<\lam_0^+<\lam_1^-<\lam_1^+<\dots <\lam_g^-<\lam_g^+.
\]
and the roots of $\Delta(\lam)=0$ by
\[
0<\lam_0<\lam_1<\dots<\lam_g.
\]
Note that $\lam_j^-<\lam_j<\lam_j^+\ (j=0,1,\dots,g)$ .

It is not easy
to calculate the asymptotic behaviour of an
Abelian integral on a general Riemann surface. 
However, as we shall see below,
the problem of describing the asymptotic behaviour of $C$
can be reduced to that of the degenerate case
\[
C_0: \mu^2=\Delta(\lam)^2.
\]

By definition, it follows that 
\begin{equation}\label{san}
\Delta(\lam)^2-4m^2=(\lam-\lam_0^-)(\lam-\lam_0^+)\cdots
(\lam-\lam_g^-)(\lam-\lam_g^+),
\end{equation}
\begin{equation}\label{yon}
\Delta(\lam)=(\lam-\lam_0)\cdots(\lam-\lam_g).
\end{equation}
Clearly, equation $(\ref{san})$ can be decomposed
\begin{equation}\label{plus}
\Delta(\lam)+2m=\prod_{j=0}^g {(\lam-\lam_j^{\sigma(j)})}
\end{equation}
and
\begin{equation}\label{minus}
\Delta(\lam)-2m=\prod_{j=0}^g {(\lam-\lam_j^{-\sigma(j)})}
\end{equation}
where $\sigma(j)=
\left\{
\begin{array}{ll}
+ & (j=g-1,g-3,\dots)  \\
- & (j=g,g-2,g-4,\dots)
\end{array}
\right.
$
and $-\sigma(j)$ denotes the opposite sign to $\sigma(j)$.\\
By (\ref{yon}), (\ref{plus}) and (\ref{minus}), we have
\begin{equation}\label{tenkai}
\lam_j-\lam_j^{\sigma(j)}=
\frac{-2m}{\prod_{k\neq j}{(\lam_j-\lam_k^{\sigma(k)})}},
\end{equation}
\begin{equation}\label{tenkai2}
\lam_j-\lam_j^{-\sigma(j)}=
\frac{2m}{\prod_{k\neq j}{(\lam_j-\lam_k^{-\sigma(k)})}}.
\end{equation}
\begin{lemma}\label{resserection}
It follows that
\[
\lam_j\udsim\lam_j^\pm.
\]
\end{lemma}
\proof Let $u_0$ be the constant term of the polynomial $\Delta(\lam)$ (see
Def.\ref{definitionofU}). 
Since $Q_1+Q_2+\cdots+Q_N<L/2$ (see section \ref{pBBS}), we obtain
\[
u_0\udsim (u_0\pm 2m),
\]
from lemma \ref{asym}, proposition \ref{zero}, and (\ref{emu}).
The proof then follows
from (\ref{san}), (\ref{yon}), and lemma \ref{poly}.$\qed$

As a corollary of lemma \ref{poly}, lemma \ref{box-Young}, and lemma \ref{resserection},
we obtain theorem \ref{renkon}. 

The
following lemmas and propositions are used to calculate the ultradiscrete limit
which is expressed as the difference of $\lam_j$s.
\begin{prop}\label{utusu}
Under the condition
\begin{equation}
j\neq k\ \Rightarrow\ U_{j+1}-U_j\neq U_{k+1}-U_k,\label{ulight}
\end{equation}
the ultradiscrete limit of $\zet{\lam_j-\lam_k^\pm}$ satisfies
\[
\left\{
\begin{array}{rcl}
\zet{\lam_j-\lam_k^\pm}\ud U_{j+1}-U_j & (j>k),& \mbox{ and }\\
\zet{\lam_j-\lam_k^\pm}\ud U_{k+1}-U_{k}  & (j<k),
\end{array}
\right.
\]
where $U_j\,(j=0,1,\dots,g)$ is the real number defined by Def.\ref{definitionofU}.
\end{prop}
\proof  From lemma \ref{poly} and lemma \ref{resserection}, it follows that
\begin{equation}\label{nemuiyou}
\lam_j\ud U_{j+1}-U_j,\quad \mbox{and}\quad \lam_k^\pm\ud U_{k+1}-U_k.
\end{equation}
The assertion is proved immediately by virture of lemma \ref{asym} and (\ref{ulight}).
$\qed$
\begin{rem}\label{a-cup}
The claim of proposition \ref{utusu} is also true
without the condition (\ref{ulight}). We can prove this assertion by using the fact
that $U_{j+1}-U_j\,(j=0,1,\cdots,g)$ can be perturbed independently
over the real numbers, as
we can naturally extend the domain of the initial condition
of the pBBS
$\{Q_n^0,W_n^0\}\subset \NN \subset \RR_{>0}$.
Continuity of the time evolution rule of pBBS ((\ref{hako}),(\ref{tama}),(\ref{kei})) justifies
the argument using such small perturbations.
\end{rem}

\section{Ultradiscretization of the solution of pd Toda equation}\label{usi}
\subsection{ultradiscretization of Abelian integrals}\label{sub-4.1}

It is not easy to describe the normalized holomorphic differential of a
Riemann surface in general case.
However, it is easy in the case of the degenerate curve $C_0$.
In fact, the normalized holomorphic differential of $C_0$ 
is expressed as
\[
\omega_j^0=\frac{1}{2\pi \rmi}\left\{
\frac{1}{\lambda-\lam_j}-\frac{1}{\lambda-\lam_0} \right\} 
d\lambda, \quad (j=1,2,\dots,g).
\]
\begin{thm} \label{Fay}
It follows that
\[
\int_{b_i}{\omega_j}\sim \int_{b_i}{\omega_j^0},\quad(\ep\to 0).
\]
\end{thm}
To prove theorem \ref{Fay}, we prepare several lemmas.
Let $\{\tilde{\omega}_j^0\}_{j=1}^g$ be holomorphic differentials
on $C$ defined by
\[
\tilde{\omega}_j^0:=\frac{1}{2\pi \rmi}\frac{\Delta(\lam)}{\sqrt{\Delta(\lam)^2-4m^2}}
\left\{\frac{1}{\lam-\lam_j}-\frac{1}{\lam-\lam_0}
\right\}\,d\lam.
\]
In the first place, let us prove:
\begin{lemma} 
It follows that
\begin{equation}\label{first}
\int_{b_i}{\tilde{\omega}_j^0}\sim\int_{b_i}{\omega_j^0}.
\end{equation}
\end{lemma}
\proof Let $X_i,\,(i=0,1,\dots,g-1)$ be real numbers which satisfy
\begin{equation}\label{X}
\lam_i^+<X_i<\lam_{i+1}^-,\quad   
\left\{
\begin{array}{ccc}
(X_i-\lam_i^+) \!\!&\!\! \udsim \!\!&\!\! X_i,\, \\
(\lam_{i+1}^--X_i) \!\!&\!\! \udsim \!\!&\!\! \lam_{i+1}^-,
\end{array}
\right.
\end{equation}
for example $X_i:=\sqrt{\lam_i^+\lam_{i+1}^-}$.
Then we obtain
\begin{eqnarray}
\; \int_{\lam_i^+}^{X_i}{\tilde{\omega}_i^0}&=&
\frac{1}{2\pi \rmi}\int_{\lam_i^+}^{X_i}{\frac{\Delta(\lam)}{\sqrt{\Delta(\lam)^2-4m^2}}
\left\{\frac{1}{\lam-\lam_i}-\frac{1}{\lam-\lam_0}\right\}d\lam}\\
&=&\frac{1}{2\pi \rmi}
\left\{\int_{\lam_i^+}^{X_i}{\frac{\prod_{k\neq i}{(\lam-\lam_k)}}{\sqrt{\Delta(\lam)^2-4m^2}}
\,d\lam}
-\int_{\lam_i^+}^{X_i}{\frac{\prod_{k\neq 0}{(\lam-\lam_k)}}{\sqrt{\Delta(\lam)^2-4m^2}}
\,d\lam}
\right\}.\label{abesi}
\end{eqnarray}
Under the condition $\lam_i^+<\lam<X_i$, 
\begin{eqnarray}
\; 
\zet{\frac{\prod_{k\neq i}{(\lam-\lam_k)}}
{\sqrt{\prod_{k\neq i}{(\lam-\lam_k^-)(\lam-\lam_k^+)}}}}&=&
\zet{\sqrt{\prod_{k\neq i}{\frac{\lam-\lam_k}{\lam-\lam_k^-}\cdot
\frac{\lam-\lam_k}{\lam-\lam_k^+}}
}}\\
&=&\zet{\sqrt{\prod_{k\neq i}{
\left\{
1+\frac{\lam_k^--\lam_k}{\lam-\lam_k^-}
\right\}
\cdot
\left\{
1+\frac{\lam_k^+-\lam_k}{\lam-\lam_k^+}
\right\}
}}}\\
&\udsim&\zet{\sqrt{\prod_{k\neq i}{
\left\{
1+\frac{\lam_k^--\lam_k}{\lam_i-\lam_k}
\right\}
\cdot
\left\{
1+\frac{\lam_k^+-\lam_k}{\lam_i-\lam_k}
\right\}}}}\\
&\udsim&
\zet{\sqrt{\prod_{k\neq i}{
\left\{
1+e^{-I_k^-/\ep}
\right\}
\cdot
\left\{
1+e^{-I_k^+/\ep}
\right\}}}},
\end{eqnarray}
where 
\begin{equation}\label{uhi}
I_k^+,I_k^->0.
\end{equation}
(The existence of positive numbers $I_k^+,I_k^-$ in
 (\ref{uhi}) is proved from proposition \ref{zero}, proposition \ref{naka}
and (\ref{emu}),
and the fact 
\[
Q_0^0+Q_1^0+\dots+Q_g^0 <\frac{L}{2}.
\]
For example, we can show
\begin{eqnarray*}
\zet{\frac{\lam_k^--\lam_k}{\lam_i-\lam_k}}&<&
\frac{2m}{\lam_0\lam_1\cdots\lam_g}\quad ,(\ep<\!\!<1)\\
&\ud&-\frac{L}{2}+(Q_0^0+Q_1^0+\dots+Q_g^0)\\
&<&0,
\end{eqnarray*}
which assures the existence of $I_k^->0$.)

Then there exist a positive number $B'>0$ such that
\begin{eqnarray}\; 
{(\ref{abesi})}
&\phantom{\sim}&\hspace{-40pt}\sim
\frac{1+O(e^{-B'/\ep})}{2\pi \rmi}{\int_{\lam_i^+}^{X_i}{\left\{\frac{1}
{\sqrt{(\lam-\lam_i^-)(\lam-\lam_i^+)}}-\frac{1}{\sqrt{(\lam-\lam_0^-)(\lam-\lam_0^+)}}\right\}
d\lam}}\nonumber\\
&\phantom{\sim}&\hspace{-40pt}
\sim\frac{1+O(e^{-B'/\ep})}{2\pi \rmi}
\left\{
2\log{\left[\sqrt{\frac{{X_i}-\lam_i^+}{\lam_i^+-\lam_i^-}}+\sqrt{\frac{{X_i}-\lam_i^-}
{\lam_i^+-\lam_i^-}}
\right]}
\right.\nonumber\\
&&\hspace{-20pt}
\left.
-2\log{\left[\sqrt{\frac{{X_i}-\lam_0^+}{\lam_0^+-\lam_0^-}}\!\!+\!\!\sqrt{\frac{{X_i}-\lam_0^-}
{\lam_0^+-\lam_0^-}}
\right]}
\!+\!2\log{\left[\sqrt{\frac{\lam_i^+-\lam_0^+}{\lam_0^+-\lam_0^-}}
\!\!+\!\!\sqrt{\frac{\lam_i^+-\lam_0^-}
{\lam_0^+-\lam_0^-}}
\right]
}\right\}\nonumber\\
&\phantom{\sim}&\hspace{-40pt}\sim
\frac{1}{2\pi \rmi}{\log{\frac{{X_i}-\lam_i}{{X_i}-\lam_0}\frac{\lam_i^+-\lam_0}{\lam_i^+-\lam_i}}}
=\int_{\lam_i^+}^{X_i}{\omega_i^0}.
\label{pin}
\end{eqnarray}
In the case of $i\neq j$, it is easy to show
\begin{equation}\label{pon}
\int_{\lam_i^+}^{{X_i}}{\tilde{\omega}_j^0}\sim\int_{\lam_i^+}^{{X_i}}{{\omega}_j^0}.
\end{equation} 
In a similar manner, it follows that
\begin{equation}\label{pan}
\int_{{X_i}}^{\lam_i^-}{\tilde{\omega_j}^0}\sim\int_{X_i}^{\lam_i^-}{\omega_j^0}.
\end{equation}
Equation (\ref{pin}), (\ref{pon}), and (\ref{pan}) complete the proof.
$\qed$

Next, we prove the following lemma, which completes the proof of theorem \ref{Fay}:
\begin{lemma}
It follows that
\[\int_{b_i}{\omega_j}\sim \int_{b_i}{\tilde{\omega}_j^0}.\]
\end{lemma}
\proof Since
$\omega_j^0$ is a holomorphic differential on $C$, the Riemann bilinear equation \cite{Munford}
gives
\[
\sum_{i=1}^g{(A_{j'i}B_{ji}^0-A_{ji}^0B_{j'i})}=0,
\]
where 
\[
A_{ji}^0=\int_{a_i}{\tilde{\omega}_j^0},\ B_{ji}^0=\int_{b_i}{\tilde{\omega}_j^0},\
A_{j'i}=\int_{a_i}{\omega_{j'}},\ B_{j'i}=\int_{b_i}{\omega_{j'}}.
\]
Then we obtain
\begin{equation}\label{firstdif}
B_{jj'}^0=\sum_{i=1}^g{A_{ji}^0B_{j'i}}.
\end{equation}
If $i\neq j$,
\begin{eqnarray}
\; 
\zet{A_{ij}^0}&=&
\zet{\frac{2}{2\pi \rmi}\int_{\lam_i^-}^{\lam_i^+}{\left\{\frac{\prod_{k\neq j}{(\lam-\lam_k)}}
{\sqrt{\Delta(\lam)^2-4m^2}}
-\frac{\prod_{k\neq 0}{(\lam-\lam_k)}}
{\sqrt{\Delta(\lam)^2-4m^2}}\right\}}d\lam}\\
&<&\frac{1+O(e^{-B''/\ep})}{\pi}
\left\{\zet{
\int_{\lam_i^-}^{\lam_i^+}{\frac{(\lam-\lam_i)d\lam}{\sqrt{(\lam-\lam_i^-)(\lam-\lam_i^+)
(\lam-\lam_k^-)(\lam-\lam_k^+)}}}}\right.\nonumber\\
&&\phantom{ooooooo}+
\left.\zet{
\int_{\lam_i^-}^{\lam_i^+}{\frac{(\lam-\lam_i)d\lam}{\sqrt{(\lam-\lam_i^-)(\lam-\lam_i^+)
(\lam-\lam_0^-)(\lam-\lam_0^+)}}}}
\right\}\\
&\udsim&\frac{\zet{\lam_i^+-\lam_i^-}}{\zet{\lam_i-\lam_k}+\zet{\lam_i-\lam_0}}\\
&\udsim&e^{-B'''/\ep},\label{seconddif}
\end{eqnarray}
where 
\begin{equation}\label{chome}
B''>0,\quad \mbox{and}\quad B'''>0
\end{equation} 
The relation (\ref{chome}) can be shown in a way similar to (\ref{abesi}).
In the case of $i=j$,
\begin{eqnarray}
A_{ii}^0&=&\frac{1}{\pi \rmi}\int_{\lam_i^-}^{\lam_i^+}{\left\{\frac{\prod_{k\neq i}{(\lam-\lam_k)}}
{\sqrt{\Delta(\lam)^2-4m^2}}
-\frac{\prod_{k\neq 0}{(\lam-\lam_k)}}
{\sqrt{\Delta(\lam)^2-4m^2}}\right\}}d\lam\\
&=&\frac{1+O(e^{-F/\ep})}{\pi \rmi}\int_{\lam_i^-}^{\lam_i^+}{\frac{d\lam}{\sqrt
{(\lam-\lam_i^-)(\lam-\lam_i^+)}}}\sim 1\label{thirddif},
\end{eqnarray}
where $F>0$.
Substituting (\ref{seconddif}) and (\ref{thirddif}) to
(\ref{firstdif}), then
\[
B_{jj'}^0 \sim B_{j'j}=B_{jj'}.
\]
$\qed$

By theorem \ref{Fay}, we can calculate the 
asymptotic behaviour in the limit $\ep\to 0$ of the period matrix 
$B=(B_{ij})_{1\leq i,j\leq g} $ associated with
the hyperelliptic curve $C : \mu^2=\Delta(\lam)^2-4m^2$. 
\begin{eqnarray}
B_{ij} \sim \int_{b_i}{\omega_j^0}
&=& 2\int_{\lam_0^+}^{\lam_i^-}
{\frac{1}{2\pi \rmi} \left\{\frac{1}{\lam-\lam_j}-\frac{1}{\lam-\lam_0}\right\}d\lam
}\\
&=&\frac{1}{\pi \rmi}\left[\log{\frac{\lam_i^--\lam_j}{\lam_i^--\lam_0}\cdot
\frac{\lam_0^+-\lam_0}{\lam_0^+-\lam_j}}\right].\label{bmat}
\end{eqnarray}
Other parameters can be calculated similarly: 
\begin{eqnarray}
\nu_j\sim\int_{\infty^+}^0{\!\!\omega_j^0} &=& \int_{-\infty}^0
{\frac{1}{2\pi \rmi} \left\{\frac{1}{\lam-\lam_j}-\frac{1}{\lam-\lam_0}\right\}d\lam}\\
&=& \frac{1}{2\pi \rmi}\,{\log{\frac{\lam_j}{\lam_0}}},
\end{eqnarray}
\begin{equation}
(\vect{r})_j\sim
\int_{\infty^+}^{\infty^-}{\!\!\!\!\omega_j^0}=2\int_{-\infty}^{\lam_0^-}
{\!\!\omega_j^0}=\frac{1}{\pi \rmi}\log{\frac{\lam_0^--\lam_j}{\lam_0^--\lam_0}},
\end{equation}
\begin{eqnarray}
\int_{\lam_0^+}^{\infty^+}{\!\!\omega_j^0}=
-\int_{-\infty}^{\lam_0^+}{\!\!\omega_j^0}=
-\frac{1}{2\pi \rmi}{\,\log{\frac{\lam_0^+-\lam_j}{\lam_0^+-\lam_0}}},
\end{eqnarray}
\begin{eqnarray}
\sum_{i=1}^g\!
{\int_{\lam_0^+}^{\mu_i}{\!\omega_j^0}}\!+\!k_j&=&
\sum_{i=1}^g
{\left\{\int_{\lam_0^+}^{\lam_{i}^-}
+\int_{\lam_{i}^-}^{\mu_i}\right\}\omega_j^0}+k_j\\
&=&\sum_{i=1}^g
{\left\{ \frac{1}{2}\int_{b_{i}} 
+\int_{\lam_{i}^-}^{\mu_i}+\frac{1}{2}\sum_{l=1}^{i-1}\int_{a_l}
\right\}\omega_j^0}+k_j.\label{muzu}
\end{eqnarray}
Using the formula for the Riemann constant corresponding to the hyperelliptic curve 
\begin{equation}
k_j=-\frac{1}{2}\sum_{i=1}^g{B_{ji}}+\frac{g+1-j}{2},
\end{equation}
(\ref{muzu}) becomes 
\begin{eqnarray}
\sum_{i=1}^g\!
{\int_{\lam_0^+}^{\mu_i}{\!\!\omega_j^0}}+k_j &=& 
\sum_{i=1}^g{\left(-\int_{\mu_i}^{\lam_i^-}{\!\!\omega_j^0}+\frac{1}{2}
\right)}\nonumber\\
\!
\!\!\!&=&\!\!\! \sum_{i=1}^g{\left[-\frac{1}{2\pi \rmi}\int_{\mu_i}^{\lam_i^-}
{\!\! \left\{\frac{1}{\lam-\lam_j}\!-\!\frac{1}{\lam-\lam_0}\right\}d\lam}
+\frac{1}{2}\right]
}\\
\!\!\!&=&\!\!\! \sum_{i=1}^g{\left[-\frac{1}{2\pi \rmi}
\log{\frac{\lam_i^--\lam_j}{\lam_i^--\lam_0}\cdot \frac{\mu_i-\lam_0}{\mu_i-\lam_j}}
+\frac{1}{2}\right]}.
\end{eqnarray}

Now, using remark \ref{oh}, proposition \ref{utusu}, and (\ref{tenkai}),
\begin{eqnarray}
\;  B_{jj}\sim\frac{1}{\pi \rmi}
\log{\frac{4m^2}{(\lam_j-\lam_0)^2 \, 
\prod_{k\geq 1}{\!(\lam_k-\lam_0)}\, \prod_{0\leq k<j}{\!(\lam_j-\lam_k)}\, 
\prod_{j<k\leq g}{\!(\lam_k-\lam_j)}}}\nonumber\\
\; \hspace{1cm}\sim \frac{1}{\pi \rmi}\frac{1}{\ep}
(2M-2(U_{j+1}\!-\!U_{j})
\!-\!\sum_{k\geq 1}{(U_{k+1}\!-\!U_{k})}\!-\!j(U_{j+1}\!-\!U_{j})\!
-\!\!\!\!\!\sum_{j<k\leq g}
{\!\!\!\!(U_{k+1}\!-\!U_{k})})\nonumber\\
\; \hspace{1cm}=\frac{1}{\pi \rmi}\frac{1}{\ep}
\left(2M-(j+1)U_{j+1}+(j+2)U_{j}+U_1{\mbox{\large\strut}}\right),\label{bii1}
\end{eqnarray}
and for $i>j$
\begin{eqnarray}
B_{ij}&\sim& \frac{1}{\pi \rmi}\log{\frac{2m(\lam_i-\lam_j)}
{(\lam_j-\lam_0)(\lam_i-\lam_0)\prod_{k\geq 1}{(\lam_k-\lam_0)}}}\\
&\sim& \frac{1}{\pi \rmi}\frac{1}{\ep}(M-(U_{j+1}-U_{j})
-\sum_{k\geq 1}{(U_{k+1}-U_{k})})\\
&=& \frac{1}{\pi \rmi}\frac{1}{\ep}
\left(M-U_{j+1}+U_{j}+U_1\hasira\right)\label{bii2}
\end{eqnarray}
where $m\ud M(=-L/2)$, and $U_{g+1}=0$.

Similarly, 
\begin{eqnarray}
\nu_j\sim\frac{1}{2\pi \rmi}\frac{1}{\ep}
\left(\hasira U_{j+1}-U_{j}-(U_1-U_0)\right),\label{defnu}\\
(\vect{r})_j\sim -\frac{1}{\pi \rmi}\frac{1}{\ep}
\left(\hasira M-U_{j+1}+U_j+U_1\right),\label{bii3}
\end{eqnarray}
and the $j$-th elements of 
$\displaystyle\vect{c}_0=\int_{\mu_0}^{\infty^+}{\vect{\!\!\!\!\omega}}
-\sum_{j=1}^g{\int_{\mu_0}^{\mu_j}{\vect{\omega}}}-\vect{K}$
\begin{eqnarray}
c_{0j}&\sim& \sum_{i=1}^g{
\left[
-\frac{1}{2\pi \rmi}\log{
\frac{(\lam_0^+-\lam_j)(\lam_i^--\lam_0)(\mu_i-\lam_j)}{ 
(\lam_0^+-\lam_0)(\lam_i^--\lam_j)(\mu_i-\lam_0)}}+\frac{1}{2}\right]}\\
&=& \sum_{i=1}^g\left[ -\frac{1}{2\pi \rmi}
\log{\frac{(\lam_j-\lam_0^+)(\lam_i^--\lam_0)(\mu_i-\lam_j)}{ 
(\lam_0^+-\lam_0)(\lam_i^--\lam_j)(\mu_i-\lam_0)}}
\right]\label{c0}
\end{eqnarray}
where we have chosen the branch $\ \log{(-1)}=\pi \rmi$. 
Using the fact that $\mu_j\,(j=1,2,\dots,g)$ are the roots of $y_{N+1}(\lam)=0$,
we can immediately calculate
the ultradiscrete limit of all the terms in (\ref{c0}) except $(\mu_i-\lam_j)$. 
(Note that $\mu_1\ud P_1-P_0$, \dots, $\mu_k\ud P_{k}-P_{k-1}$, \dots,
$\mu_{g}\ud -P_{g-1}.$)
Indeed,
\begin{eqnarray*}
\prod_{i=1}^g{(\lam_i^--\lam_0)}\ud \sum_{i=1}^g{(U_{i+1}-U_{i})}=-U_1,
\end{eqnarray*} 
\begin{eqnarray*}
\zet{\prod_{i=1}^g{\!\!(\lam_i^--\lam_j)}}&\udsim& \prod_{1\leq i<j}{\!\!(\lam_j-\lam_i)}\times
(\lam_j-\lam_j^-)\times\prod_{j<i\leq g}{\!\!(\lam_i-\lam_j)}\\
&=&\prod_{1\leq i<j}{\!\!(\lam_j-\lam_i)}\times
{\frac{2m}{ \prod_{l\neq j}{\zet{\lam_l-\lam_j}}}}\times\prod_{j<i\leq g}{\!\!(\lam_i-\lam_j)}\\
\\
&\ud& M-(U_1-U_0),
\end{eqnarray*}
and
\begin{eqnarray*}
\prod_{i=1}^g{(\mu_i-\lam_0)} &\udsim& \prod_{i=1}^g{\mu_i}\\
&\ud& \sum_{1\leq i\leq g}{\!\!(P_i-P_{i-1})}\\
&=& -P_0.
\end{eqnarray*}

Unfortunately, it is not easy
to calculate the terms $\displaystyle\prod_{i=1}^g{(\mu_i-\lam_j)}$.
For the time being, 
we treat these terms formally as
\[\prod_{i=1}^g{(\mu_i-\lam_j)}\ud \Xi_j.\]
We will prove a concrete expression for $\Xi_j$ in the Appendix.
Thus, we obtain
\begin{equation}
c_{0j}\sim\frac{1}{2\pi \rmi}\frac{1}{\ep}
\left(\,g(M-U_{j+1}+U_j+U_0)-P_0-\Xi_j\hasira\right).
\end{equation}

The fundamental decomposition of $B$ is given as
\begin{eqnarray}
-\frac{1}{2\pi \rmi}\frac{1}{\ep}
\left(\begin{array}{@{\,}cccc@{\,}}
	A_{1} & 0 & \ldots & 0 \\
	0 & A_{2} & \ldots & 0 \\
	\vdots & \vdots & \ddots & \vdots \\
	0 & 0 & \ldots & A_{g} 
\end{array}\right)
=\Gamma_g\dots \Gamma_3\Gamma_2 B \Gamma_2^t\Gamma_3^t\dots \Gamma_g^t
\end{eqnarray}
where 
\begin{eqnarray*}
\hspace{-5pt}
A_{k}=\frac{k+1}{k}(-M-(k+1)U_{k}+kU_{k+1}),\quad (k=1,2,\dots,g,\ U_{g+1}=0).
\end{eqnarray*}
and
\begin{eqnarray*}
\Gamma_2=
\left(\begin{array}{@{\,}ccccc@{\,}}
1    & 0 & 0 & \ldots & 0 \\
-1/2 & 1 & 0 & \ldots &  0 \\
-1/2 & 0 & 1 & \ldots & 0 \\
\vdots & \vdots & \vdots & \ddots & \vdots \\
-1/2 & 0 & 0 & \ldots & 1 
\end{array}\right),
\Gamma_3=
\left(\begin{array}{@{\,}ccccc@{\,}}
	1 & 0 & 0       & \ldots & 0 \\
	0 & 1 &  0      & \ldots & 0 \\
	0 & -1/3 &   1     & \ldots & 0  \\
	\vdots & \vdots & \vdots & \ddots & \vdots \\
	0 & -1/3 &    0    & \ldots & 1 
\end{array}\right),\\
,\dots,
\Gamma_g=
\left(\begin{array}{@{\,}ccccc@{\,}}
	1 & 0 & \ldots & 0 & 0 \\
	0 & 1 & \ldots & 0 & 0 \\
\vdots & \vdots & \ddots & \vdots & \vdots \\
	0 & 0 & \ldots & 1 & 0 \\
	0 & 0 & \ldots & -1/g & 1 
\end{array}\right).
\end{eqnarray*}
Thus, using (\ref{defnu}), we obtain
\begin{eqnarray}\label{periodic}
B^{-1}\vect{\nu}=\frac{1}{2}(\varsigma_1,
\varsigma_2,\dots,\varsigma_g)^t
\end{eqnarray}
where
\[
\begin{array}{ccr}
\varsigma_1\!\!\!\!\!&=&\!\!\!\!\!
\displaystyle-\frac{1}{2}\frac{b_1}{c_1}+\frac{1}{2\cdot 3}\bunsuu{2}
+\frac{1}{3\cdot 4}\bunsuu{3}+\cdots+\frac{1}{g(g+1)}\bunsuu{g},\\
\varsigma_2\!\!\!\!\!&=&\!\!\!\!\!\displaystyle
\phantom{-\frac{1}{2}\frac{b_1}{a_1}}\ 
-\frac{1}{3}\bunsuu{2}+\frac{1}{3\cdot 4}\bunsuu{3}+\cdots+
\frac{1}{g(g+1)}\bunsuu{g},\\
\varsigma_3\!\!\!\!\!&=&\!\!\!\!\!
\displaystyle-\frac{1}{4}\bunsuu{3}+\cdots+\frac{1}{g(g+1)}\bunsuu{g},\\
&&\vdots\\
\varsigma_g\!\!\!\!\!&=&\!\!\!\!\!\displaystyle-\frac{1}{g+1}\bunsuu{g},
\end{array}
\]
with
\[
b_k=U_0-(k+1)U_k+kU_{k+1},\ c_k=-M-(k+1)U_k+kU_{k+1}.
\]
On the other hand, (\ref{bii1}), (\ref{bii2}), and (\ref{bii3})
yield the important relation
\begin{eqnarray}
\vect{r}&=&-\frac{1}{g+1}(\vect{b}_1+\vect{b}_2+\cdots+\vect{b}_g)\\
&=&-\frac{1}{N}(\vect{b}_1+\vect{b}_2+\cdots+\vect{b}_g),\label{syuukisei}
\end{eqnarray}
where the $\vect{b}_j$ is $j$-th column vector of the period matrix $B$.
\subsection{Ultradiscretization of the theta function
solution to the pd Toda}\label{ultraman}
In this subsection, 
we will calculate the ultradiscrete limit 
of the meromorphic function of the form
\begin{equation}\label{mero}
\; 
\Psi_j(\vect{z}):=
\hen{z_j}\log{\frac{\theta(\vect{z},B)}{\theta(\vect{z}-\frac{1}{N}
(\vect{b}_1+\cdots+\vect{b}_g),B)}},\ \ \ \mbox{where  } \vect{z}=
(z_1,\dots,z_g)^t
\end{equation}
rather than the theta function
itself, because we want to ultradiscretize
(\ref{solutionofToda}) with (\ref{syuukisei}).
We introduce the real matrix $B^\circ$ and the real vector
 $\vect{z}^\circ$ by
$
B=i\BBB,\ \ \vect{z}=i\BBB\Zzz
$.
Starting from Def.\ref{theta},
\begin{eqnarray}
\theta(\vect{z},B)&=&\sum_{\vect{n}\in\ZZ^g}{\exp{(\pi \rmi\vect{n}^t B\vect{n}+2\pi \rmi
\vect{n}^t\vect{z})}}\\
&=& \sum_{\vect{n}\in\ZZ^g}{\exp{(-\pi \vect{n}^t \BBB\vect{n}-2\pi 
\vect{n}^t \BBB\Zzz)}}\\
&=& \sum_{\vect{n}\in\ZZ^g}{\exp{(-\pi(\vect{n}+\Zzz)^t\BBB(\vect{n}+\Zzz))}
\exp{(\pi{\Zzz}^t\BBB\Zzz)}},
\end{eqnarray}
and
\begin{eqnarray}
\; 
\hen{z_j}\theta(\vect{z},B)&=& 2\pi \rmi\sum_{\vect{n}\in\ZZ^g}{n_j 
\exp{(\pi \rmi\vect{n}^t B\vect{n}+2\pi \rmi
\vect{n}^t\vect{z})}}\\
&=&
2\pi \rmi\sum_{\vect{n}\in\ZZ^g}{\!\!\!
n_j \exp{(-\pi(\vect{n}+\Zzz)^t\BBB(\vect{n}+\Zzz))}
\exp{(\pi{\Zzz}^t\BBB\Zzz)}}.
\end{eqnarray}
Using these formulae, (\ref{mero}) becomes
\begin{eqnarray}
\; 
\Psi_j(\vect{z})&\!\!\!=&
\frac{\theta_j(\Zzz)\theta(\Zzz-\vect{e})-
\theta(\Zzz)\theta_j(\Zzz-\vect{e})
}{\theta(\Zzz)\theta(\Zzz-\vect{e})}\\
&\!\!\!=& \!\!2\pi \rmi \frac
{
\sum_{\vect{n},\vect{m}}
{(n_j\!-\!m_j)\exp{(-H(\vect{n}+\!\Zzz))} \exp{(-H(\vect{m}+\Zzz\!-\vect{e}))}}}
{
\sum_
{\vect{n},\vect{m}}{\exp{(-H(\vect{n}+\Zzz))} \exp{(-H(\vect{m}+\Zzz-\vect{e}))}}}
\label{merobibun}
\end{eqnarray}
where $\vect{e}=(1/N,1/N,\dots,1/N)^t$, and $H(\vect{x})=
\vect{x}^t (\pi \BBB)\vect{x} $.

Since $B\sim O(\ep^{-1})$, we define $\Gamma(\vect{x}):=
\lim_{\ep\to +0}{\ep H(\vect{x})}$.  
In (\ref{merobibun}), it turns out that
the ultradiscrete behaviour
of $\Psi_j(\vect{z})$ is strongly dependent on
the term $(n_j\!-\!m_j)$.
Recalling the fact that the period matrix must satisfy
$\mathrm{Im}{B}\geq 0$, 
we find that
$\Gamma(\vect{x})\geq 0,$ $\forall \vect{x}\in\RR^g$.
Since $\Gamma$ is a quadratic form over $\RR^g$,
we can order all the elements of $\ZZ^g \times\ZZ^g$ as
\begin{eqnarray*}\; 
0\leq \Gamma(\vect{z}^\circ+\vect{n}^{(1)})+
\Gamma(\vect{z}^\circ+\vect{m}^{(1)}-\vect{e})
\leq \Gamma(\vect{z}^\circ+\vect{n}^{(2)})
+\Gamma(\vect{z}^\circ+\vect{m}^{(2)}-\vect{e})
\\ 
\hspace{115pt}\leq \Gamma(\vect{z}^\circ+\vect{n}^{(3)})
+\Gamma(\vect{z}^\circ+\vect{m}^{(3)}-\vect{e})\leq\dots,
\end{eqnarray*}
$((\vect{n}^{(i)},\,\vect{m}^{(i)})
\in\ZZ^g\times\ZZ^g,\ i=1,2,\dots)$.

Let $n^{(i)}_k$ and $m^{(j)}_k$ be $k$-th element of $\vect{n}^{(i)}$ and
$\vect{m}^{(j)}$. 
Then, the asymptotic behaviour of $\Psi_j(\vect{z})$ is described as
\begin{eqnarray}
\; 
\Psi_j(\vect{z})\sim 2\pi \rmi \sum_{i=1}^{\infty}{(n_j^{(i)}-
m_j^{(i)})}\exp{\left[\frac{-1}{\ep}
\left(\Gamma(\vect{n}^{(i)}+\vect{z}^\circ)
+\Gamma(\vect{m}^{(i)}+\vect{z}^\circ-\vect{e})\right.\right.}\nonumber\\
\hspace{3.3cm}\left.\left.\phantom{\frac{1}{\ep}}
-\Gamma(\vect{n}^{(1)}+\vect{z}^\circ)
-\Gamma(\vect{m}^{(1)}+\vect{z}^\circ-\vect{e})\right)\right].\label{pko}
\end{eqnarray}
Let \[\;  G_{i}(\vect{z}):=
\Gamma(\vect{n}^{(i)}+\vect{z}^\circ)
+\Gamma(\vect{m}^{(i)}+\vect{z}^\circ-\vect{e})
-\Gamma(\vect{n}^{(1)}+\vect{z}^\circ)
-\Gamma(\vect{m}^{(1)}+\vect{z}^\circ-\vect{e}).\]
Since $\Gamma$ is a positive definite quadratic form over $\RR^g$,
the set $\{\vect{x}\in\RR^g\,\vert\,\zet{\vect{x}}=R\}\subset \RR^g$
is bounded for any $R>0$. Thus, the set
\[
\{i\in\NN\,\vert\,\Gamma(\vect{n}^{(i)}+\vect{z}^\circ)
+\Gamma(\vect{m}^{(i)}+\vect{z}^\circ)=R
\}
\]
is a finite set for any $R$. 
We arrange all the elements of $\{G_{i}(\vect{z})\}_{i\in\NN}$ as
\begin{eqnarray*}
0\leq G_{1}(\vect{z})=\dots=
 G_{\sigma(1)}(\vect{z})
<G_{\sigma(1)+1}
(\vect{z})=\dots= G_{\sigma(2)}(\vect{z})<\dots.
\end{eqnarray*}
The relation (\ref{pko}) becomes 
\begin{eqnarray}
\Psi_j(\vect{z})&\sim& 2\pi \rmi\sum_{p=1}^{\infty}
{\sum_{l=\sigma(p-1)+1}^{\sigma(p)}{(n_j^{(l)}-m_j^{(l)})}
\exp{\left[-\frac{G_{l}(\vect{z})}{\ep}\right]}}\\
&=&2\pi \rmi\sum_{p=1}^{\infty}
{
\sum_{l=\sigma(p-1)+1}^{\sigma(p)}{(n_j^{(l)}-m_j^{(l)})}
\exp{\left[-\frac{G_{l}(\vect{z})}{\ep}\right]}
}\\
&=&2\pi \rmi\sum_{p=1}^{\infty}
{
I_p
\exp{\left[-\frac{{G_{\sigma(p)}}(\vect{z})}{\ep}\right]}}
,
\end{eqnarray}
where $I_p=\sum_{l=\sigma(p-1)+1}^{\sigma(p)}{(n_j^{(l)}-m_j^{(l)})}$.

Let
\[
q(j):=\min{\{p\in\NN\,\vert\, I_p\neq 0\}}.
\]

To calculate the ultradiscrete limit of (\ref{solutionofToda}), 
we recall the calculations in section \ref{sub-4.1}, and notice the following
relation,
\[
\int_{a_j}{\lam\omega_j}\sim\, \lam_j-\lam_0,
\]
which is obtained from 
$\displaystyle
\omega_j\sim \omega_j^0=\frac{1}{2\pi \rmi}\left\{\frac{1}{\lam-\lam_j}
-\frac{1}{\lam-\lam_0}
\right\}\,d\lam.
$
Hence, the coefficient $c_{j,g-1}$ defined in Sec,\ref{ne}
is found to be:
\[
c_{j,g-1}=\frac{\lam_j-\lam_0}{2\pi \rmi}.
\]
Substituting these relations in (\ref{solutionofToda}), we obtain
\begin{eqnarray}
I_{n+2}^t\!+\!V_{n+1}^t&=&\!\sum_{i=0}^g{(I_i^0+V_i^0)}-\sum_{i=0}^g{(\lam_i-\lam_0)}
-\sum_{j=1}^g{\!\frac{\lam_j-\lam_0}{2\pi \rmi}\Psi_j(\vect{z})}\\
&=& \sum_{i=0}^g{\lam_i}-\sum_{i=0}^g{(\lam_i-\lam_0)}
-\sum_{j=1}^g{\frac{\lam_j-\lam_0}{2\pi \rmi}\Psi_j(\vect{z})}\\
&=&(g+1)\lam_0-\sum_{j=1}^g{\frac{\lam_j-\lam_0}{2\pi \rmi}\Psi_j(\vect{z})},\label{saisyuuteki}
\end{eqnarray}
where $\vect{z}=n\vect{m}+t\vect{\nu}+\vect{c}(0)$.

The following formula gives an answer to the initial value problem of pBBS.
\begin{thm}
The ultradiscretization of (\ref{saisyuuteki}) is given by
\begin{equation}\label{main}
\min{[Q_{n+2}^t,W_{n+1}^t]}=
\min_j{  \left[ 
U_0-U_1,\,
(U_j-U_{j+1})+\tilde{G}_{q(j)}(\vect{z})
\right]},
\end{equation}
where $\tilde{G}_{q(j)}(\vect{z}):=
\displaystyle\lim_{\ep\to0}{G_{q(j)}(\vect{z})}$ with
$\vect{z}=n\vect{m}+t\vect{\nu}+\vect{c}(0)$.
\end{thm}
\proof By lemma \ref{asym}, (\ref{main}) is obvious
when $U_0-U_1$, and $(U_j-U_{j+1})+\tilde{G}_{q(j)}(\vect{z})
,\, (j=1,2,\dots,g)$
are all distinct.
In the general case, we have only to consider small perturbations
as in remark \ref{a-cup}.
Since we can make $U_0-U_1$ and $(U_j-U_{j+1})+\tilde{G}_{q(j)}(\vect{z})$
all distinct by perturbing $U_j-U_{j+1}$ independently,
we can conclude that (\ref{main}) holds in 
the generl case by continuity of both sides
of the equation.
$\qed$

\begin{rem}\label{syuuki}
Note that 
$\Psi_j(n\vect{m}+t\vect{\nu}+\vect{c}(0))$ does not change
under the translation
\[
t\vect{\nu}\mapsto t\vect{\nu}+\vect{b}_k,\quad \forall k
\]
or equivalently
\[
t(B^{-1})_k\vect{\nu}\mapsto t(B^{-1})_k\vect{\nu}+1,
\]
where $(B^{-1})_k$ is the $k$-th column of $B^{-1}$.
\end{rem}

\section{Fundamental cycles of the periodic box-ball systems}\label{tora}
\subsection{relative period}
Let $\bbs{N}$ be a set of $N$-soliton states with no 0-soliton.
We treat separately the set of $N$-soliton
states with 0-solitons, which is denoted by $\bbs{N}^*$.
A state $x\in\bbs{N}$
is expressed as 
\[
x=x_1x_2\dots x_L\ \ \ \ \ \mbox{for}\ \ x_i\in\{0,1\}.
\]
Using the translation map 
\[
\mathrm{S}\,:\, \bbs{N}\to\bbs{N}\ \ \ \ \ x_1x_2\dots x_L
\mapsto x_2\dots x_Lx_1
\]
which sends the first letter to the last, we define the set 
$\mathcal{T}(x) \subset \bbs{N}$ (for \,$x\in\bbs{N}$),
\[
\mathcal{T}(x):=\{y\in\bbs{N}\,\vert\, \exists m \in\ZZ_{\geq 0}\ \ 
\mathrm{s.t.} \ \ \mathrm{S}^m(x)=y\}.
\]
Let us denote the 10-elimination by $\mathrm{El}\,:\,
\bbs{N}\to\bbs{N}\cup\bigcup_{n<N}{\bbs{n}^\ast}
, $ and the time evolution in the pBBS by
$T\,:\, \bbs{N}\to\bbs{N}$. We also define $V\,:\,\bbs{N}\cup
\bbs{N}^*\to\bbs{N}$ as the map which acts as the
identity on $\bbs{N}$ and eliminates the
0-solitons in $\bbs{N}^*$.
\begin{rem}
$\mathrm{El}$ is bijective.  $V\circ\mathrm{El}$ is surjective, 
but not injective. 
\end{rem}
\begin{defi}
The fundamental cycle $f(x)$ of $x\in\bbs{N}(\mbox{or}\in\bbs{N}^*
)$ is the minimum positive integer $p$
that satisfies $T^p(x)=x$.
The relative period of $x\in\bbs{N}(\mbox{or}\in\bbs{N}^*
)$, $r(x)$, is the minimum positive integer $q$
for which $T^q(x)\in\mathcal{T}(x)$.
\end{defi}
\begin{rem}
$r(x)\,\vert\, f(x)$ for any $x\in\bbs{N}$ because $x\in\mathcal{T}(x)$.
\end{rem}
\begin{rem}\label{choi}
$\mathrm{S}$ and $T$  commute. And $\mathrm{S}$ and $\el$ also 
commute. Consequently,\[
\mathcal{T}(x)=\mathcal{T}(y)\Leftrightarrow
\mathcal{T}(\el(x))=\mathcal{T}(\el(y)).
\]
\end{rem}
Since the method we presented in section \ref{ne} can
only be used to calculate the
relative period of pBBS systems, the following claims are
important.
\begin{lemma}\label{commute}
Let $x\in\bbs{N}$ be an $N$-soliton state of the
pBBS. It holds that
\[
\mathcal{T}(\mathrm{El} \circ T^n(x))=\mathcal{T}(T^n\circ \mathrm{El}(x)),
\quad n\in\NN
\]
or
\[
\mathcal{T}(\mathrm{El}^{-1}\circ T^n(x))=\mathcal{T}(T^n\circ \mathrm{El}^{-1}
(x)).
\]
\end{lemma}
\proof Let $Q_n^t$ and $W_n^t$ $(n=1,2,\dots
,N,\, t\in\NN )$ be numbers defined by an $N$-soliton
state $x$ (see section \ref{pBBS} figure \ref{naming2}).
Note that the equations (\ref{hako})-(\ref{tamakazu}) give
$Q_n^{t+1}$ and $W_n^{t+1}$ from $Q_n^t$ and $W_n^t$.
Obviously, when we replace $Q_n^t$ and $W_n^t$ to $Q_n^t-1$ and $W_n^t-1$,
then these equations give $Q_n^{t+1}-1$ and $W_n^{t+1}-1$,
which means 
\[
\mathcal{T}(\el\circ T(x))=\mathcal{T}(T\circ \el(x)).
\]
Using this formula, the former assertion is easily
proved by induction. 
To prove the latter assertion, we start from the formar assertion.
\begin{eqnarray}
\; 
\mathcal{T}(\mathrm{El} \circ T^n(x))=\mathcal{T}(T^n\circ \mathrm{El}(x))
&\Leftrightarrow& \exists m>0\ \mbox{s.t.}\ \el\circ T^n(x)
=\mathrm{S}^m \circ
T^n\circ\el(x)\nonumber\\
&\Leftrightarrow& \el\circ T^n\circ\mathrm{El}^{-1}(y)=\mathrm{S}^m
\circ T^n(y)\nonumber\\
 &\Leftrightarrow& T^n\circ\mathrm{El}^{-1}(y)=\mathrm{El}^{-1}
\circ\mathrm{S}^m\circ T^n(y)\label{this}
\end{eqnarray}
where $y=\el (x)$.
Recalling remark \ref{choi}, (\ref{this}) is equivalent to
$  T^n\circ\mathrm{El}^{-1}(y)=\mathrm{S}^m\circ\mathrm{El}^{-1}\circ T^n(y)$,
which completes the proof. $\qed$
\begin{prop}\label{rel}
If a state $x\in\bbs{N}$ satisfies the condition $\mathrm{El}(x)\in\bbs{N-1}^*$,
then $r(x)=f(\mathrm{El}(x))$.
\end{prop}
\proof Let $\tilde{x}=\mathrm{El}(x)$.
Since $\mathrm{El}(x)\in\bbs{N-1}^*$, $\tilde{x}$ has exactly one 0-soliton.
If $T^p(\tilde{x})=\tilde{x}$, 
\[\; 
\mathcal{T}(x)=\mathcal{T}(\el^{-1}(\tilde{x}))
=\mathcal{T}(\el^{-1}\circ T^p(\tilde{x}))
=\mathcal{T}(T^p\circ\el^{-1} (\tilde{x}))
=\mathcal{T}(T^p(x)).
\]
So, $r(x)\leq f(\tilde{x})$. Conversely, if 
$\mathcal{T}(x)=\mathcal{T}(T^q(x))
$, it follows that
\[
\mathcal{T}(\tilde{x})=\mathcal{T}(\el(x))=\mathcal{T}(\el\circ T^q(x))=
\mathcal{T}(T^q\circ \el(x))=\mathcal{T}(T^q(\tilde{x})).
\]
Since a 0-soliton does not move under the time evolution, 
the fact that $\tilde{x}$ has exactly one 0-soliton leads to
$\tilde{x}=T^q(\tilde{x})$. So, $r(x)\geq f(\tilde{x})$.$\qed$

\begin{rem}\label{sym}
In the proof of proposition \ref{rel} , we conclude that $\mathcal{T}(\tilde{x})
=\mathcal{T}(T^q(\tilde{x})) \Rightarrow \tilde{x}=T^q(\tilde{x})$.
This claim fails in the case where there are more than two 0-solitons
in the sequence of $\el(x)$,
where are arranged symmetrically. In this situation, 
\[
r(x)\lvertneqq f(\tilde{x}).
\]
 We call this symmetry `internal symmetry'.
Internal symmetry makes the problem more complicated.
We do not concider this symmetry in the present paper.
\end{rem}
The statement of proposition \ref{rel} can be generalized as follows.
\begin{cor}\label{Nagosi}
If a state $x\in\bbs{N}$ satisfies the condition $\mathrm{El}(x)\in\bbs{n}^*$, where
$n<N$, and $x$ is without internal symmetry,
then $r(x)=f(\mathrm{El}(x))$.
\end{cor}
By virtue of Cor.\ref{Nagosi}, the fundamental period of the pBBS can be 
obtained from the relative period of the corresponding pBBS.
\subsection{Formula for the fundamental period}
Recall the definition of the Young diagram associated with the state of pBBS
(Section \ref{paiotu}).
We also define
\[
n_l:=\{\mbox{the number of the rows of which length is $L_l$}\},
\]
\[\hspace{-9pt}
l_0:=L-\sum_{j=1}^l{2p_j},\ \ \ \ l_j:=L_j-L_{j+1},\ \ \ \ N_j:=l_0+\sum_{l=1}^j
{2n_l(L_l-L_{j+1})}.
\]

The fundamental cycle of $x\in\bbs{N}$ can be described by using the 
data of the corresponding Young diagram.
In fact, the following formula gives the fundamental cycle of the pBBS system.
\begin{thm}\label{yura}
Let $x\in\bbs{N}$ be a $N$-soliton pBBS without internal symmetry. Then 
\[
f(x)=\lcm{\left(\frac{N_sN_{s-1}}{l_sl_0},\frac{N_{s-1}N_{s-2}}{l_{s-1}l_0},
\dots, \frac{N_1N_0}{l_1l_0},1\right)}.
\]
\end{thm}
\begin{rem}\label{reduce}
Recalling proposition \ref{rel}, the formula in proposition \ref{yura} is equivalent to
\[
r(x)=\lcm{\left(\frac{N_{s-1}N_{s-2}}{l_{s-1}l_0},
\dots, \frac{N_1N_0}{l_1l_0},1\right)}
\]
because one can obtain the Young diagram corresponding to $\mathrm{V}\circ
\mathrm{El}(x)$ by eliminating the first column of the Young diagram
corresponding to $x$.
\end{rem}
Though this formula was first obtained by elementary combinatorial methods
\cite{Yoshihara}, 
we can obtain the same formula using a different method 
relying on the results of the previous sections.

From (\ref{solutionofToda}), (\ref{periodic}), and remark \ref{syuuki}, the relative period
$r(x)$ satisfies
\begin{equation}\label{aaru}
r(x)=
\lcm{\left(\frac{2}{\varsigma_1},
\frac{2}{\varsigma_2},\dots,\frac{2}{\varsigma_g},1\right)},
\end{equation}
where $\varsigma_j$ are numbers defined by (\ref{periodic}).
We use the following lemmas
to prove theorem \ref{yura}.

This lemma claims that there is a simple relation between
the ultradiscrete limit of Riemann surfaces and Young diagrams.
The proof of this lemma is given in the Appendix.
\begin{cor}\label{elieli}
$\varsigma_k=\varsigma_{k+1}\Leftrightarrow$
the lengths of
$k$-th and $(k+1)$-th rows from the bottom in the Young diagram are equal.
\end{cor}
\proof \begin{eqnarray*}
\; 
\varsigma_k=\varsigma_{k+1}&\Leftrightarrow&
 -\frac{1}{k+1}\frac{b_k}{c_k}+\frac{1}{(k+1)(k+2)}\frac{b_{k+1}}{c_{k+1}}
=-\frac{1}{k+2}\frac{b_{k+1}}{c_{k+1}}\\
&\Leftrightarrow& \frac{b_k}{c_k}=\frac{b_{k+1}}{c_{k+1}}\\
&\Leftrightarrow& \frac{U_0-(k+1)U_k+kU_{k+1}}{-M-(k+1)U_{k}+kU_{k+1}}
\!\!=\!\!\frac{U_0-(k+2)U_{k+1}+(k+1)U_{k+2}}{-M-(k+2)U_{k+1}+(k+1)U_{k+2}}\\
&\Leftrightarrow& U_k-U_{k+1}=U_{k+1}-U_{k+2}.
\end{eqnarray*}
Lemma \ref{box-Young} completes the proof.$\qed$

The following lemma is almost trivial. We omit the proof.
\begin{lemma}\label{suu}
Let $p,p',p'',q,q',q''$ be integers, and $p,p',p''$ are relatively
prime to $q,q',q''$ respectively.
If \[
\frac{q}{p}-\frac{q'}{p'}=\frac{q''}{p''},
\]
then
$
\lcm{(q,q')}=\lcm{(q,q'')}.
$
\end{lemma}
\noindent \underline{Proof of theorem \ref{yura}}\quad
First, we prove the theorem for the case where 
\begin{equation}\label{kinniku}
i\neq j\Rightarrow\varsigma_i\neq\varsigma_j.
\end{equation}
Starting from (\ref{aaru}),
\begin{eqnarray}
\;  r(x)&\!\!=&\!\!\lcm{
\left(\frac{2}{\varsigma_1},\dots,\frac{2}{\varsigma_g},1\right)}\\
&\!\!=&\!\!\lcm{\left(\frac{2}{\varsigma_1+\sum_{j=1}^g{\varsigma_j}},
\frac{2}{\varsigma_1-\varsigma_2},
\dots,\frac{2}{\varsigma_k-\varsigma_{k+1}},\dots,
\frac{2}{\varsigma_{g-1}-\varsigma_g},1
\right)}\label{motteiku}
\end{eqnarray}
by lemma \ref{suu}. From (\ref{periodic}),
\begin{eqnarray*}
\; 
\varsigma_k-\varsigma_{k+1}&=&-\frac{1}{k+1}\frac{b_k}{c_k}+
\left(\frac{1}{(k+1)(k+2)}-\frac{1}{k+2}\right)\frac{b_{k+1}}{c_{k+1}}
\nonumber\\
&=& \dots \nonumber\\
&=& \frac{(-M-U_0)(U_{k+2}-2U_{k+1}+U_k)}{(\!-M-(k+1)U_k+kU_{k+1})
(\!-M-(k+2)U_{k+1}+(k+1)U_{k+2})}.
\end{eqnarray*}
On the other hand,
by definition of $l_j,N_j$ and lemma \ref{box-Young}, we obtain
\[
l_j=L_{j+1}-L_j=U_j-U_{j+1}-(U_{j-1}-U_j)=-U_{j+1}+2U_j-U_{j-1},\]
\[l_0=L-2U_0,\]
and
\begin{eqnarray*}
N_j&=&l_0+2\sum_{l=1}^j{n_l(L_l-L_{j+1})}\\
&=& L-2(j+1)U_{j}+2jU_{j+1}\quad(\because (\ref{kinniku})
\stackrel{\mathrm{Cor}.\ref{elieli}}{\Longrightarrow} n_j=1).
\end{eqnarray*}
Recalling $M=-L/2$, we obtain
\[\varsigma_{j}-\varsigma_{j+1}=
\frac{2l_0l_{j+1}}{N_{j+1}N_{j}},\quad (j=1,2,\dots,g-1).
\]
And, by definition of $\varsigma_j$ ((\ref{periodic})), we derive
\[
\varsigma_1+\sum_{j=1}^g{\varsigma_j}=-\frac{b_1}{c_1}=
-\frac{U_0-2U_1+U_2}{-M-2U_1+U_2}=\frac{2l_1}{N_1}
(=\frac{2l_1l_0}{N_1N_0}).
\]
By (\ref{motteiku}), it follows that
\[
r(x)=\lcm{\left(\frac{N_gN_{g-1}}{l_gl_0},\dots,\frac{N_1N_0}{l_1l_0},1
\right)}.
\]
The fact that $\,(\ref{kinniku})\Rightarrow s=N=g+1\,$
completes the proof under the condition (\ref{kinniku}).

For general cases, let us define $\varrho(i)\, (i=1,2,\dots,s)$ by
\[\; \varsigma_1=\dots=\varsigma_{\varrho(1)}>
\varsigma_{\varrho(1)+1}=\dots
=\varsigma_{\varrho(2)}>\varsigma_{\varrho(2)+1}=\dots>
\varsigma_{\varrho(s-1)-1}=\dots=\varsigma_{\varrho(s)},
\]
and $\varrho(0):=0$.
We obtain
\[
l_j=-U_{\varrho(j+1)}+2U_{\varrho(j)}-U_{\varrho(j-1)},
\]
and
\begin{eqnarray*}
N_j&=&l_0+2\sum_{l=1}^j{n_l(L_l-L_{\varrho(j+1)})}\\
&=&l_0+2\sum_{k=1}^s{
\sum_{\varrho(k-1)+1\leq l\leq\varrho(k)}
{(\varrho(k)-\varrho(k-1))(L_l-L_{\varrho(j+1)})}}\\
&=& L-2(\varrho(j)+1)U_{\varrho(j)}+2\varrho(j)U_{\varrho(j)+1}.
\end{eqnarray*}
We can complete the proof
in a similar manner to that of the previous case. $\qed$

\bigskip
\noindent {\it Acknowledgment.}
The authors are very grateful to Professor Ralph Willox for helpful comments
on this paper.
\appendix
\section{The proofs of the remained lemmas}
\subsection{Proof of lemma \ref{box-Young}}
\begin{figure}[htbp]
\includegraphics[viewport=102 554 257 708, clip,
width=4cm,height=4cm]{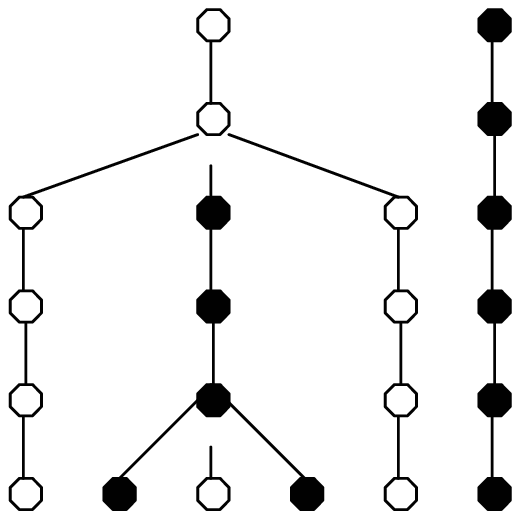}
\begin{picture}(100,100)(-30,0)
\put(0,0){11111000100111111000000000000}
\put(0,20){11110001111100000000000}
\put(0,40){1110011110000000000}
\put(0,60){110111000000000}
\put(0,80){11100000000}
\put(0,100){110000000}
\put(-150,0){$5$}
\put(-150,20){$4$}
\put(-150,40){$3$}
\put(-150,60){$2$}
\put(-130,0){$3$}
\put(-110,0){$1$}
\put(-104,16){$*$}
\put(-90,0){$2$}
\put(-70,0){$6$}
\put(-113,23){$3$}
\put(-110,40){$2$}
\put(-110,60){$1$}
\put(-104,76){$*$}
\put(-70,20){$5$}
\put(-70,40){$4$}
\put(-70,60){$3$}
\put(-113,83){$3$}
\put(-110,100){$2$}
\put(-55,0){$12$}
\put(-55,20){$11$}
\put(-55,40){$10$}
\put(-50,60){$9$}
\put(-50,80){$8$}
\put(-50,100){$7$}
\put(60,10){$\uparrow\mbox{10-elimination}$}
\put(50,30){$\uparrow$}
\put(40,50){$\uparrow$}
\put(30,70){$\uparrow$}
\put(25,90){$\uparrow$}
\put(23,110){$\vdots$}
\put(-103,111){$\vdots$}
\put(-43,111){$\vdots$}
\end{picture}
\caption{A state of pBBS with $N=3$ and the associated graph.
The nodes at the bottom the graph are associated with 
$(Q_1,W_1,Q_2,W_2,Q_3,W_3)=(5,3,1,2,6,12)$.}
\label{graph1}
\end{figure}

\begin{figure}[htbp]
\includegraphics[viewport=90 600 252 700, clip,
width=4cm,height=2.6cm]{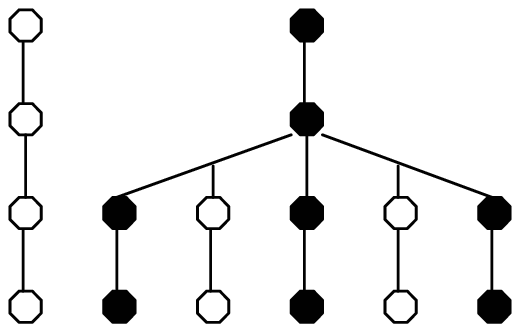}
\begin{picture}(100,70)(-31,0)
\put(0,0){1111111000110000110000000}
\put(0,20){1111110010001000000}
\put(0,40){1111100000000}
\put(0,60){1111000000}
\put(-150,0){$7$}
\put(-150,20){$6$}
\put(-150,40){$5$}
\put(-150,60){$4$}
\put(-130,0){$3$}
\put(-130,20){$2$}
\put(-110,0){$2$}
\put(-110,20){$1$}
\put(-92,0){$4$}
\put(-92,20){$3$}
\put(-94,42){$8$}
\put(-92,60){$7$}
\put(-72,0){$2$}
\put(-72,20){$1$}
\put(-53,0){$7$}
\put(-53,20){$6$}
\put(-80,56){$t_1$}
\put(-101,10){$t_2$}
\put(-63,10){$t_3$}
\put(-135,50){$t_4$}
\put(-77,42){$P$}
\put(60,10){$\uparrow\mbox{10-elimination}$}
\put(50,30){$\uparrow$}
\put(40,50){$\uparrow$}
\put(30,70){$\vdots$}
\put(-84,71){$\vdots$}
\put(-142,71){$\vdots$}
\put(-104,30){$*$}
\put(-67,30){$*$}
\end{picture}\vspace{6mm}\\
\hspace{2mm}
\includegraphics[viewport=113 580 209 677, clip,
width=2.6cm,height=2.5cm]{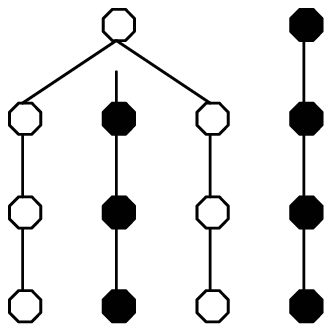}
\begin{picture}(100,70)(-62,0)
\put(0,0){11111100011100000}
\put(0,20){1111100110000}
\put(0,40){111101000}
\put(0,60){11100}
\put(50,10){$\uparrow$}
\put(30,30){$\uparrow$}
\put(15,50){$\uparrow$}
\put(10,70){$\vdots$}
\put(-143,0){$6$}
\put(-143,20){$5$}
\put(-143,40){$4$}
\put(-122,0){$3$}
\put(-122,20){$2$}
\put(-122,40){$1$}
\put(-101,0){$3$}
\put(-101,20){$2$}
\put(-101,40){$1$}
\put(-80,0){$5$}
\put(-80,20){$4$}
\put(-80,40){$3$}
\put(-80,60){$2$}
\put(-123,62){$3$}
\put(-73,71){$\vdots$}
\put(-114,71){$\vdots$}
\put(-116,55){$\ast$}
\end{picture}
\caption{Examples of the graph associated with two typical states.
In the first case, two blocks of 0's disappear simultaneously.
In the second case, two adjacent blocks disappear simultaneously,
which would require writing the $\ast$ twice.
We shall write it only once.}
\label{graph3}
\end{figure}

To prove lemma \ref{box-Young}, we investigate the 10-elimination 
(Section \ref{paiotu}) in detail.
Let us consider a state consisting of $N$ blocks of consecutive 1's and 
0's, which are arranged alternatingly.
We denote the length of the $k$-th block of consecutive 1's by $Q_k$
and $k$-th consecutive 0's by $W_k$.
In order to show how these blocks are reconstructed by 10-eliminations,
it is convenient to draw a graph which consists of nodes and links.
For example, let us consider a state where $N=3$, $Q_1=5$,
$W_1=3$, $Q_2=1$, $W_2=2$, $Q_3=6$, and $W_3=12$.
(See figure \ref{graph1}.)
The blocks of 1's are represented by white nodes, and those of
0's by black nodes. A number is associated with
each node. The numbers in the bottom of the graph are equal
to $Q_1,W_1,\dots,Q_N,W_N$ respectively.
Going up we arrange by one step, these numbers decrese by $1$.
The sign `$\ast$' means zero, and a
 blocks of consecutive 1's or 0's 
disappears at the point where $\ast$ appears.
When one block disappears, the two blocks adjacent to a `$*$'
join together. 
Figure \ref{graph3} show examples of
the graphs associated with a typical state.

Let us define several terms relating to
this associated graph.
\begin{defi}
A tree is a connected component in the associated graph.
\end{defi}
Note that any tree has exactly one $\ast$.
\begin{rem}
Only two types of tree can exist. One is a tree consisting
of white nodes, and the other is a tree of black nodes.
We denote the `white tree' as a`w-tree', and 
`black tree' as a `b-tree'. 
\end{rem}
\begin{defi}\label{A.2}
Let $P$ be a node, and $t$ be 
a tree in the associated graph.\\
The height of $P$, denoted by $\mathrm{Ht}(P)$, is the number of
links in the path from $P$ to the bottom of the graph. \\
And the height of $t$, denoted by $\mathrm{Ht}(t)$, is the height of $\ast$
contained within $t$.
\end{defi}

Let us denote by $\Phi_x$ the graph associated with the state $x\in\bbs{N}$.
We introduce a semiordering on the set of trees by
\[
t,t'\in\Phi_x,\ t\succ t'\,\Leftrightarrow\, t\ \mbox{ straddles } t'.
\]
Now we define two important sets.
\begin{defi}
Let $t\in\Phi_x$ be a tree, then we define a set of trees as
\[
\mathrm{Und}(t):=\{s\in\Phi_x: \mbox{tree}\,\vert\, s\preceq t
\},
\]
and a set of integers as
\[
\mathrm{Ft}(t):=\left\{A_{\sigma(i)}\in X\,
\left\vert\, 
\begin{array}{c}
A_{\sigma(i)}
\mbox{ is an associated number with}\\ 
\mbox{a node at the foot of tree $t$}.
\end{array}
\right\}\right. ,
\]
with $X=\{A_{\sigma(i)}\}_{i=1}^{2N}$, $A_{2l-1}=Q_l$, and $A_{2l}=W_l$.
\end{defi}
\begin{figure}[htbp]
\begin{center}
\includegraphics[viewport=102 554 257 708, clip,
width=4cm,height=4cm]{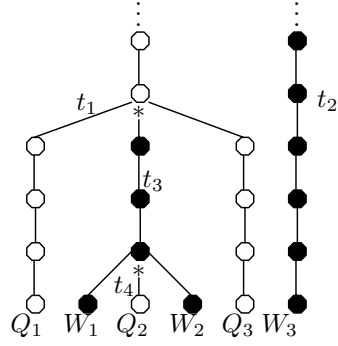}
\begin{picture}(100,100)(-30,0)
\put(-150,-5){$Q_1$}
\put(-130,-5){$W_1$}
\put(-110,-5){$Q_2$}
\put(-104,16){$*$}
\put(-90,-5){$W_2$}
\put(-70,-5){$Q_3$}
\put(-55,-5){$W_3$}
\put(-104,76){$\ast$}
\put(-103,111){$\vdots$}
\put(-43,111){$\vdots$}
\put(-125,79){$t_1$}
\put(-34,80){$t_2$}
\put(-100,50){$t_3$}
\put(-111,10){$t_4$}
\end{picture}
\end{center}
\caption{An example clarifying Def.\ref{A.2}}
\label{graph5}
\end{figure}

\noindent \underline{Example}
In figure \ref{graph5}, $t_1$ and $t_4$ are white trees, and $t_2$ and $t_3$
are black trees. The trees have
the relation 
$
t_4\prec t_3\prec t_1
$,\,
$\und(t_1)=\{t_3,t_4\}$, 
$\ft(t_1)=\{Q_1,Q_3\}$, etc.
The height of $t_3$ and $t_4$ are given as
 $\mathrm{Ht}(t_3)=4$, and $\mathrm{Ht}(t_4)=1$.
Note that 
\begin{equation}\label{white}
 t \mbox{ is a w-tree }\Leftrightarrow \ \ft(t)\subset \{Q_i\}_{i=1}^N,
\end{equation}
and
\begin{equation}\label{black}
t \mbox{ is a b-tree }\Leftrightarrow \ \ft(t)\subset \{W_i\}_{i=1}^N.
\end{equation}

\begin{lemma}\label{omake}
Let $t\in\Phi_x$ be a tree. The number of links in $t$
is equal to 
$\displaystyle
\sum_{A_{\sigma(i)}\in\ft(t)}{\!\!\!A_{\sigma(i)}}
$.
\end{lemma}
\proof This is a natural consequence of the definition of the associated graph.$\qed$

Remember that a
maximal element $\xi$ in a semiorderd set $X$ is the element which satisfies
\[
\xi'\in X,\ \ \xi\preceq \xi'\,\Rightarrow\, \xi=\xi'.
\]
\begin{rem}
Let $t\in\Phi_x$ be a w-tree, and $s\in\Phi_x$ be a b-tree.
Then any maximal element of $\und(t)\setminus \{t\}$
is a b-tree, and any maximal element of $\und(s)\setminus \{t\}$
is a w-tree.
\end{rem}
We call the node which is connected with more than $2$ links 
the branch point.
We define
the multiplicity of the branch point $P$ as 
\[
m_P:=\{\mbox{the number of links connected to $P$}\}-2.
\]
Note that 
\[
\{\mbox{the number of maximal element in }\und(t)\!\setminus\! \{t\}\}
=\sum_{P:\mbox{branch pt. in $t$}}{\hspace{-21pt}m_P}.
\]
\noindent \underline{Example}
In figure \ref{graph3}, the multiplicity $m_P$ is equal to $2$.
The maximal elements of $\und(t)\,\setminus\,\{t\}$ are
$t_2$ and $t_3$. Note that $\mathrm{Ht}(P)=\mathrm{Ht}(t_2)=\mathrm{Ht}(t_3)=2$.

Let us define
\[
\ft(\und(t)):=\bigcup_{s\in\und(t)}{\ft(s)}, 
\]
where $t\in\Phi_x$ is a tree.
\begin{lemma}\label{tree}
Let $t\in\Phi_x$ be a w-tree, and $s\in\Phi_x$ be a b-tree.
Then, 
\begin{equation}\label{whole1}
\sum_{t'\in\mathrm{Und}(t)}{\!\!\!\mathrm{Ht}(t')}=\sum_{Q_i
\in\ft(\und(t))\cap \{Q_j\}}{\!\!\!\!\!Q_i},
\end{equation}
and 
\begin{equation}\label{whole2}
\sum_{s'\in\mathrm{Und}(s)}{\!\!\!\mathrm{Ht}(s')}=\sum_{W_i
\in\mathrm{Ft}(\mathrm{Und}(s))\cap \{W_j\}}{\!\!\!\!\!W_i}.
\end{equation}
\end{lemma} 
\proof Let $h:=\#\und(t)$. We prove the lemma by induction of $h$.

When $h=1$, a w-tree $t$ is a straight segment of length $Q_j$, 
where $\{Q_j\}=\und(t)$.
Then, $\mathrm{Ht}(t)=Q_j$.
Hence, recalling (\ref{white}), we can conclude that (\ref{whole1}) is true 
immediately. Similarly, (\ref{whole2}) is true for a b-tree $s$.

Suppose that (\ref{whole1}) and (\ref{whole2}) hold for $h=1,2,\cdots,p$.
Let $t$ be a w-tree with $\#\und(t)=p+1$.
Denote the branch points which belong to $t$ by $P_1,P_2,\dots,P_r$.
\begin{figure}[htbp]
\begin{center}
\includegraphics[viewport=144 541 406 727, clip,
width=4cm,height=4cm]{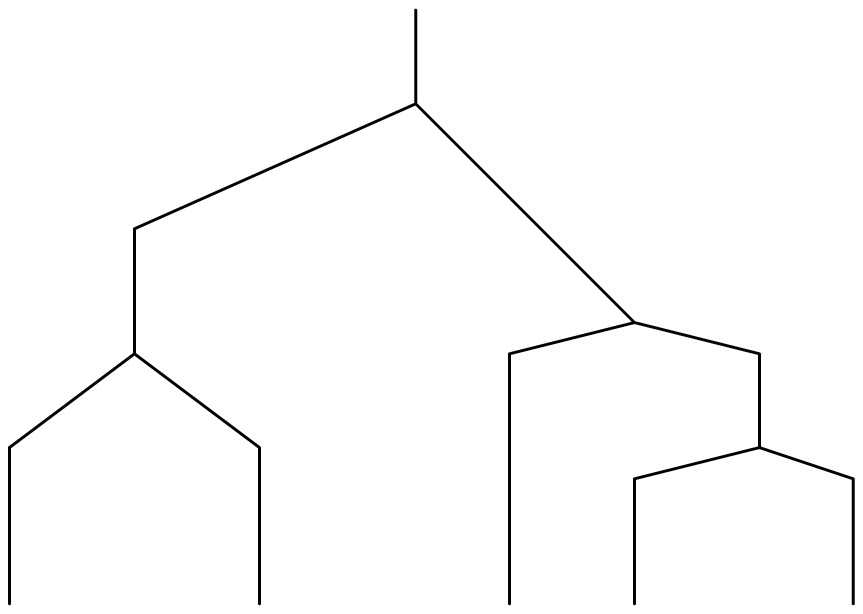}
\begin{picture}(0,90)(120,-4)
\put(0,0){\vector(0,1){105}}
\put(0,5){\vector(0,-1){5}}
\put(43,0){\vector(0,1){43}}
\put(43,5){\vector(0,-1){5}}
\put(120,0){\vector(0,1){50}}
\put(120,5){\vector(0,-1){5}}
\put(90,0){\vector(0,1){28}}
\put(90,5){\vector(0,-1){5}}
\put(140,0){\vector(0,1){90}}
\put(140,5){\vector(0,-1){5}}
\put(54,107){$\ast$}
\put(24,43){$P_1$}
\put(58,88){$P_2$}
\put(85,52){$P_3$}
\put(100,29){$P_4$}
\put(-25,60){Ht$(t)$}
\put(43,20){$\gamma_1$}
\put(90,10){$\gamma_4$}
\put(121,25){$\gamma_3$}
\put(141,45){$\gamma_2$}
\end{picture}
\end{center}
\caption{An example of tree $t$. $(\gamma_j:=\mathrm{Ht}(P_j))$.}
\label{tukare}
\end{figure}

From lemma \ref{omake}, we obtain
\begin{eqnarray}
\{\mbox{the number of links consisted in }t\}&=&
\sum_{Q_i\in\ft(t)}{Q_i}\nonumber\\
&=&\mathrm{Ht}(t)+\sum_{j=1}^r{\mathrm{Ht}(P_j)}.\label{uhyou}
\end{eqnarray}
(See figure \ref{tukare}.) Since we can assume that the maximal elements 
$s_1,s_2,\dots,s_r\in\und(t)\!\setminus\!\{t\}$
satisfy $\mathrm{Ht}(s_j)=\mathrm{Ht}(P_j)\ (j=1,2,\dots,r)$,
(\ref{uhyou}) becomes
\begin{equation}\label{nolo}
\sum_{Q_i\in\ft(t)}{Q_i}=\mathrm{Ht}(t)+\sum_{j=1}^r{\mathrm{Ht}(s_j)}.
\end{equation}
On the other hand, we easily find that
\begin{equation}\label{nolonolo}
\sum_{j=1}^r{\mathrm{Ht}(s_j)}=\sum_{t'\in\und(t)\setminus\{t\}}
{\mathrm{Ht}(t')}-\sum_{j=1}^r{\left(\sum_{t''\in
\und(s_j)\setminus\{s_j\}}{\mathrm{Ht}(t'')}\right)}.
\end{equation}
Let
$\tilde{t}_{jk}\ (k=1,2,\dots,l_j)$ be maximal elements of
$\und(s_j)\!\setminus\!\{s_j\}$.
By the induction hypothesis and (\ref{whole1}) we obtain
\begin{eqnarray}
\sum_{j=1}^r{\left(\sum_{t''\in
\und(s_j)\setminus\{s_j\}}{\mathrm{Ht}(t'')}\right)}
&=&\sum_{j=1}^r{\sum_{k=1}^{l_k}{\sum_{t''\in\und(\tilde{t}_{jk})}{\mathrm{Ht}(t'')}
}}\\
&=&\sum_{j=1}^r{\sum_{k=1}^{l_k}{\sum_{Q_i\in\ft(\und(\tilde{t}_{jk}))\cap\{
Q_n\}}{\!\!\!\!\!\!\!\!Q_i}
}}\\
&=&\sum_{Q_i\in(\ft(\und(t)\setminus\{t\}))}{\!\!\!\!Q_i}.\label{nolonolonolo}
\end{eqnarray}
Substituting (\ref{nolonolo}) and (\ref{nolonolonolo}) to (\ref{nolo}),
we find that (\ref{whole1}) holds for $h=p+1$.
Equation (\ref{whole2}) can be proved 
in a similar manner.$\qed$

Let \[
H_k:=\{\mbox{the height of the $k$-th smallest tree in $\Phi_x$}\}.
\]
Note that the proof of lemma \ref{box-Young} is completed by proving the following two formulae
(\ref{kaelu}) and (\ref{inu}).
\begin{equation}\label{kaelu}
\min{\left\{
\sum_{\{A_{\sigma(i)}\}\in\mathcal{B}(N-k,N)}{\!\!\!\!\!\!\!\!\!\! A_{\sigma(i)}}
\,\,\right\}}=H_1+\cdots+H_k.
\end{equation}
\begin{equation}\label{inu}
\; 
H_1+\cdots+H_k=
\left\{
\begin{array}{l}
\mbox{the number of boxes}\\
\mbox{from under to the $k$-th
step in the Young diagram}
\end{array}
\right\}.
\end{equation}
\noindent \underline{Proof of (\ref{kaelu})}\quad\  
Let $\{A_{\sigma(i)}\}^*$ be an element of $\mathcal{B}(N-k,N)$
which satisfies
\[
\min{\left\{
\sum_{\{A_{\sigma(i)}\}\in\mathcal{B}(N-k,N)}{\!\!\!\!\!\!\!\!\!\! A_{\sigma(i)}}
\,\,\right\}}
=
\sum_{\{A_{\sigma(i)}\}^*\in\mathcal{B}(N-k,N)}{\!\!\!\!\!\!\!\!\!\! A_{\sigma(i)}}.
\]
Without loss of generality, we can assume $Q_\alpha\in\{A_{\sigma(i)}\}^*$ for
some $\alpha$. Let $t\in\Phi_x$ be a tree with $Q_\alpha\in\ft(t)$.
It follows that
\begin{equation}\label{pyoin}
Q_\beta\in\ft(\und(t))\Rightarrow Q_\beta\in\{A_{\sigma(i)}\}^*.
\end{equation}
($\because$ Otherwise, there exists $Q_{\beta_1},\dots,Q_{\beta_l}$
with $Q_{\beta_j}\!\in\ft(\und(t))$
 and 
$Q_{\beta_j}\!\not\in\{A_{\sigma(i)}\}^*$.
Let $P_j$ be the nearest branch point to $Q_{\beta_j}$.
Without loss of generality,
we can assume $\mathrm{Ht}(P_1)=\min{(\mathrm{Ht}(P_j))}$.  
Let us define the subgraph
\[
\Phi_x^*:=\left\{P:\mbox{node}\,
\left\vert\,
\begin{array}{l} 
\mbox{the path }P_1\to P
\mbox{ consists of links} \\
\mbox{extending below $P_1$}
\end{array}
\right\}\right.,
\]
and 
\[
\Phi_x^{**}:=\{\mbox{the node and links straddled by $\Phi^*_x.$
}\}.
\]
Note that $\Phi_x^{**}
\setminus \{Q_{\beta_1}\} \subset \{A_{\sigma(i)}\}^*$.
The inequality \[\sum_{Q_j\in\Phi_x^{**}
\setminus \{Q_{\beta_1}\}}{Q_j}
>\sum_{W_i\in\Phi_x^{**}}{W_i}\]
 leads to a contradiction with the definition 
of $\{A_{\sigma(i)}\}^*$.
 )

The relation (\ref{pyoin}) claims
\[\; 
\{A_{\sigma(i)}\}^*=\left\{\coprod_j({\ft(\und(t_j))}\cap\{Q_n\})\right\}\amalg
\left\{\coprod_j({\ft(\und(s_m))}\cap\{W_n\})\right\}
\]
for some w-trees $t_j$ and b-trees $s_m$. From lemma \ref{tree}, we obtain
\[
\sum_{\{A_{\sigma(i)}\}^*\in\mathcal{B}(N-k,N)}{\!\!\!\!\!\!\!\!\!\! A_{\sigma(i)}}
=H_{\tau_1}+H_{\tau_2}+\cdots+H_{\tau_l},
\]
for some $\tau$.
It completes the proof of (\ref{kaelu}).$\qed$

\noindent \underline{Proof of (\ref{inu})}\quad
Note that the number of boxes below the $k$-th step in the Young
diagram is equal to the number of bridges connected to the blocks
which disappear by 10-eliminations, when the number of solitons becomes
$N-k$. (See figure \ref{0-solitons}.)
By definition of the associated graph $\Phi_x$, it is clear that 
the equation (\ref{inu}) holds. $\qed$

From proposition \ref{naka}, (\ref{kaelu}) and (\ref{inu}),
Lemma \ref{box-Young} is proved.

\subsection{Proof of proposition \ref{elem} and \ref{elem2}}\label{later}
In this subsection, we give the proof of proposition \ref{elem} and \ref{elem2}.
Recall (\ref{kisei}):
\[
\left\{
\begin{array}{c}
x_{n+1}=(\lam-a_n)x_n-b_nx_{n-1}\\
y_{n+1}=(\lam-a_n)y_n-b_ny_{n-1}
\end{array}
\right.
\]
and the initial condition 
$
\left(\begin{array}{@{\,}c@{\,}}
	x_0 \\
	x_1
\end{array}\right)=
\left(\begin{array}{@{\,}c@{\,}}
	0 \\
	1
\end{array}\right),
$
$
\left(\begin{array}{@{\,}c@{\,}}
	y_0 \\
	y_1
\end{array}\right)=
\left(\begin{array}{@{\,}c@{\,}}
	1 \\
	0
\end{array}\right).
$

Let $X_j=\lam-a_j$ and $Y_j=-b_j$. For convenience,
we denote $X_j$ by $(j)$, and $Y_j$ by $(j-1,j)$ in this section.
For example, $(a)(b-1,b)$ means $X_a\cdot Y_b$.

Let
\[\; 
\Omega_{jk}:=\left\{
\begin{array}{r}
\{i_1,i_2,\dots,i_s,j_1,j_2,\dots,j_t\}\\
\subset \ZZ/N\ZZ
\end{array}
\!
\left\vert\!
\begin{array}{c}
2s+t=k+1-j,\\
\{i_1-1,i_1,\dots,i_s-1,i_s,j_1,\dots,j_t\}\\
\hfill=\{j,j+1,\dots,j_k\}
\end{array}
\right\}\right. .
\]
We define 
\[\; 
((j,j+1,\dots,k)):=\!\!\!\!\!\!\!\!\!\!\hspace{-0.4cm}
\sum_{\!\!\!\!\!\!\!\!\!\!\{i_1,i_2,\dots,i_s,j_1,\dots,j_t\}\in\Omega_{jk}}
{\hspace{-0.9cm}(i_1-1,i_1)(i_2-1,i_2)\cdots(i_s-1,i_s)(j_1)(j_2)\cdots(j_t)}.
\]

For example, $((1,2,3))=(1)(2)(3)+(1,2)(3)+(1)(2,3)=X_1X_2X_3+
Y_2X_3+X_1Y_3$. We also defined $((2,3,\dots,k))$ by the similar rule.
(For example, $((2,3))=(2)(3)+(2,3)=X_2X_3+Y_3$.)
\begin{lemma}\label{hodai}
If $2\leq n \leq N+1$, then $x_n=((1,2,\dots,n-1))$. 
And if $3\leq n \leq N+1$, then $y_n=Y_1\times ((2,3,\dots,n-1))$.
\end{lemma}
\proof We prove the lemma by induction.
When $n=2$, $x_2=(\lam-a_1)x_1-b_1x_0=\lam-a_1=X_1=((1))$.
Since $x_{n+1}=X_nx_n+Y_nx_{n-1}$, 
from the induction hypothesis,
$x_{n+1}=(n)\cdot((1,2,\dots,n-1))+(n-1,n)\cdot
((1,2,\dots,n-2))=((1,2,\dots,n))$. Hence $x_n=((1,2,\dots,n))$ holds
for any $n$.
We can prove the assertion
for $y_n$ in a similar fashion. $\qed$

If $n>N$, $n$ should be considered as an element of $\ZZ/N\ZZ$ because
of the periodic boundary condition $a_{N+j}=a_j,\,b_{N+j}=b_j$. 
In this sense, we may write $Y_1=(N,1)$. 
Propositions \ref{elem} and \ref{elem2} are proved immediately from lemma \ref{hodai}.
In fact,
the relation $y_{N+1}=Y_1\times ((2,3,\dots,N))$ is equivalent
to proposition \ref{elem2}. In order to prove proposition \ref{elem}, note that
$\Delta(\lam)=x_{N+1}+y_N=((1,2,\dots,N))+(N,1)((2,3,\dots,N-1))$.
Let us decompose $\mathcal{A}(N)\subset 2^{\ZZ/N\ZZ}$ into two disjoint sets
\[
\mathcal{A}(N)=\mathcal{A}'\sqcup \mathcal{A}''
\]
where $\mathcal{A}'$ consists of the subsets ($\subset \ZZ/N\ZZ$) that include
 $(N,1)$, and $\mathcal{A}'':=\mathcal{A}-\mathcal{A}'$.
Proposition \ref{elem} is obtained by the fact that
\[\hspace{-12pt}
\sum_{(j_1-1,j_1,\dots,j_k-1,j_k)\in\mathcal{A}'}
{Y_{j_1}\dots Y_{j_k}X_{i_1}\dots X_{i_{N-2k}}}=(N,1)((2,3,\dots,N-1))
\]
and
\[
\sum_{(j_1-1,j_1,\dots,j_k-1,j_k)\in\mathcal{A}''}
{Y_{j_1}\dots Y_{j_k}X_{i_1}\dots X_{i_{N-2k}}}=((1,2,\dots,N)).
\]
\subsection{Calculation of $\Xi_j$}
Recall that $a_j\, (j=0,1,\dots,g)$ are the roots of $\Delta(\lam)=0$,
and $\mu_j\,(j=1,2,\dots,g)$ are the roots of $y_{N+1}(\lam)=0$.
Hence,
\[\; 
\prod_{i=1}^g{(\mu_j-a_i)}=\frac{\Delta(\mu_j)}{\mu_j-a_0}
=\frac{[((1,2,\dots,N))+(N,1)((2,3,\dots,N-1))]_{\lam=\mu_j}}{\mu_j-a_0},
\]
where $N=g+1$.\\
Since
\[((2,3,\dots,N))\vert_{\lam=\mu_j}=0\]
 and 
\[((1,2,\dots,N))=
(1)((2,3,\dots,N))+(1,2)((3,4,\dots,N)),\]
we find
\[
\prod_{i=1}^g{(\mu_j-a_i)}=\frac{[(1,2)((3,4,\dots,N))+(N,1)((
2,3,\dots,N-1))]_{\lam=\mu_j}}{\mu_j-a_0}.
\]
Note that $(j)=\lam-(I_{j+1}+V_j)$ and $(j-1,j)=-I_jV_j$. 

We show some examples here.\\
\noindent \underline{Example}\\
If $N=2\,(g=1)$, 
\begin{eqnarray*}
-(\mu_1-a_1)=\frac{I_0V_0+I_1V_1}{\mu_1-a_0}&\udsim&\frac{I_0V_0+I_1V_1}{\mu_1}\\
&\ud& -\min{\{Q_0^0+W_0^0, Q_1^0+W_1^0 \}}-(-P_1)\\
&\equiv&\Xi_1.
\end{eqnarray*}
If $N=3\,(g=2)$,
\begin{eqnarray}\label{ex}
\prod_{i=1}^2{(\mu_j-a_i)}=\frac{-I_2V_2(\mu_j-(I_1+V_0))
-I_1V_1(\mu_j-(I_0+V_2))}{\mu_j-a_0}.
\end{eqnarray}
To calculate the ultradiscrete limit of the right hand side of (\ref{ex}), we have to compare 
the magnitude of each term.\\
For example, 
when $j=1$, $(P_2-P_1)<\min{\{Q_1^0,W_0^0\}}$, and
$Q_2^0+W_2^0+P_2-P_1<Q_1^0+W_1^0+\min{\{P_2-P_1,\,Q_0^0,\,W_2^0\}}$, then
\[\Xi_1=Q_2^0+W_2^0+P_2-P_1.
\]

\end{document}